\documentclass[11pt,a4paper]{article}
\usepackage[utf8]{inputenc}
\usepackage{amsmath,amsthm,amssymb,amsfonts}
\usepackage{scalerel}
\usepackage{mathtools}
\usepackage{graphicx}
\usepackage{subfigure}
\usepackage{mathrsfs}
\usepackage{bbold}
\usepackage[]{units}
\usepackage{bm}
\usepackage{braket}
\usepackage{color}
\usepackage{makecell}
\usepackage{enumitem}
\usepackage{upgreek}
\usepackage{blindtext}
\usepackage{graphics}
\usepackage{verbatim} 

\usepackage[dvipsnames]{xcolor}
\usepackage{bbm}
\usepackage[bb=boondox]{mathalfa}
\usepackage{array}
\usepackage{multirow}
\usepackage{tabularx}
\usepackage{float}
\usepackage{graphicx}
\usepackage{dcolumn}
\usepackage{bm}
\usepackage{amsmath,amsfonts,amssymb}
\usepackage{slashed}
\usepackage{braket,xcolor}
\usepackage{verbatim}
\usepackage{multirow}
\usepackage{amsfonts, mathtools,resizegather}
\usepackage[export]{adjustbox}

\usepackage{caption,jheppub}
\DeclareMathOperator{\sech}{sech}

\DeclareMathOperator{\arcsinh}{arcsinh}

\footnotetext{Authors' names are listed in alphabetical order.}
\footnotetext{Corresponding authors.}

\title{\boldmath {Operator growth in open quantum systems: lessons from the dissipative SYK}}

\author[a]{Budhaditya Bhattacharjee,}
\author[b,*]{Xiangyu Cao,}
\author[c,*]{Pratik Nandy,}
\author[a]{and Tanay Pathak}

\affiliation[a]{Centre for High Energy Physics, Indian Institute of Science,\\C.V. Raman Avenue, Bangalore 560012, India}
\affiliation[b]{Laboratoire de Physique de l'Ecole Normale Sup\'erieure, ENS, Universit\'e PSL,\\
CNRS, Sorbonne Universit\'e, Universit\'e de Paris, 75005 Paris, France}
\affiliation[c]{Center for Gravitational Physics and Quantum Information,\\ Yukawa Institute for Theoretical Physics, Kyoto University,\\ Kitashirakawa Oiwakecho, Sakyo-ku, Kyoto 606-8502, Japan}

\emailAdd{budhadityab@iisc.ac.in}
\emailAdd{xiangyu.cao@ens.fr}
\emailAdd{pratik@yukawa.kyoto-u.ac.jp}
\emailAdd{tanaypathak@iisc.ac.in}

\abstract{We study the operator growth in open quantum systems with dephasing dissipation terms, extending the Krylov complexity formalism of \cite{Parker:2018yvk}.
Our results are based on the study of the dissipative $q$-body Sachdev-Ye-Kitaev (SYK$_q$) model, governed by the Markovian dynamics. We introduce a notion of ``operator size concentration'' which allows a diagrammatic and combinatorial proof of the asymptotic linear behavior of the two sets of Lanczos coefficients ($a_n$ and $b_n$) in the large $q$ limit. Our results corroborate with the semi-analytics in finite $q$ in the large $N$ limit, and the numerical Arnoldi iteration in finite $q$ and finite $N$ limit. As a result, Krylov complexity exhibits exponential growth following a saturation at a time that grows logarithmically with the inverse dissipation strength. The growth of complexity is suppressed compared to the closed system results, yet it upper bounds the growth of the normalized out-of-time-ordered correlator (OTOC).  We provide a plausible explanation of the results from the dual gravitational side.}

\begin{document} 
\maketitle
\flushbottom

\section{Introduction} \label{intro}

A proper understanding of operator growth in dissipative systems is of fundamental interest. This is because dissipative phenomena are ubiquitous in nature and for any practical purposes, it is desirable to understand the effect of the environment on the system itself. One can consider the system plus the environment as a whole system that follows the unitary dynamics \cite{Breuer2007}. Tracing out the environment leads the system dynamics to non-unitary. However, for the Markovian environment, the system dynamics can be efficiently formulated in terms of Lindbladian evolution \cite{Lindblad1976, Gorini}. The Lindbladian consists of two parts, a part usually known as the Hermitian Liouvillian which governs the unitary closed system dynamics, and another part which includes the information of the environment. This part is not Hermitian and thus makes the whole evolution non-unitary.

There has been a growing interest to study operator growth in systems whose dynamics are governed by such Lindbladian. Especially it is of utmost interest how the environment affects the chaotic and integrable nature of the system \cite{PhysRevLett.61.1899, PhysRevLett.123.254101, Xu:2020wky}. Several recent studies have taken promising approaches using different probes, namely, the spectral statistics \cite{Li:2021kuv, Garcia-Garcia:2021rle, Matsoukas-Roubeas:2022odk, Kawabata:2022cpr}, operator-size distribution \cite{Zhang:2022knu, Omanakuttan:2022ikz, pfz}, the out-of-time-ordered correlator (OTOC) \cite{PhysRevA.103.062214, Schuster:2022bot, Weinstein:2022yce, Syzranov:2017zyp}, and Krylov (K-)complexity \cite{Parker:2018yvk, Bhattacharya:2022gbz, Liu:2022god}. 

In this paper, we aim to study the operator growth in the dissipative Sachdev-Ye-Kitaev (SYK) model from the point of view of the K-complexity \cite{Parker:2018yvk} (for an incomplete list of studies, see \cite{Rabinovici:2020ryf, Jian:2020qpp, Barbon:2019wsy, Dymarsky:2019elm, Cao:2020zls, Dymarsky:2021bjq, Caputa:2021sib, 
Rabinovici:2021qqt,
Bhattacharjee:2022vlt, Balasubramanian:2022tpr, 
Hornedal:2022pkc,
Caputa:2021ori,
Bhattacharjee:2022ave,
Bhattacharjee:2022qjw,  Balasubramanian:2022dnj,  Rabinovici:2022beu,  He:2022ryk, Guo:2022hui} and the references therein). The SYK model is a $(0+1)$-dimensional quantum mechanical model which consists of $N$ fermions where $q$ of them are interacting at a time \cite{PhysRevLett.70.3339}. The model is integrable for $q=2$ and chaotic for $q >2$. Moreover, the model is exactly solvable, and it is particularly analytically tractable in the large $N$ and the large $q$ limit \cite{Maldacena:2016hyu}. In particular, at low temperatures, it shows an emergence of conformal symmetry which allows writing a Schwarzian action \cite{Maldacena:2016hyu}. This action shares close similarity with the effective action in AdS$_2$ gravity \cite{Maldacena:2016upp}, for which it received considerable attention from the holographic side \cite{Kittu, Cotler:2016fpe}. The chaotic nature of this model is well captured by OTOC, satisfying the chaos bound \cite{Maldacena:2015waa}. Several recent studies have also studied K-complexity in SYK analytically for closed system \cite{Parker:2018yvk, Bhattacharjee:2022ave} and numerically for closed \cite{Jian:2020qpp, Rabinovici:2020ryf} and open/dissipative systems \cite{Liu:2022god}. The numerical results are performed for finite $N$ and finite $q$ (especially $q=4$) \cite{Jian:2020qpp, Rabinovici:2020ryf, Liu:2022god}, where the Krylov bases have been constructed using the Lanczos algorithm \cite{Lanczos1950AnIM}. For the dissipative systems, one can implement the algorithms suitable to non-unitary dynamics, namely the Arnoldi iteration \cite{Arnoldipaper}, which was studied in \cite{Bhattacharya:2022gbz}. The algorithms are numerical, thus one needs to restrict to a finite number of fermions. However, to see the universal properties of the Lanczos coefficients, especially its asymptotic growth in the presence of the environment, analytical studies are highly desirable. Here we report some analytical results in the large $N$ and the large $q$ limit for the dissipative SYK which has not been reported before.

The dissipative model we choose is proposed by Kulkarni, Numasawa and Ryu \cite{Kulkarni:2021xsx} (see \cite{Sa:2021tdr} for a similar model). Recently, it has been shown that the model is dual to a non-Hermitian two-site SYK model at low temperature, connected by a Keldysh wormhole \cite{Garcia-Garcia:2022adg}.
It is closely analogous to a Maldacena-Qi \cite{Maldacena:2018lmt} coupled SYK model, albeit with some significant differences. 
To establish our results, we take several different approaches. First, we give a diagrammatic and combinatorial proof of our claim Eq.~\eqref{eq:an-main} and Eq.~\eqref{eq:bn-main} for the SYK model. The proof relies on the $1/q$ expansion, introducing the notion of ``operator size concentration''. Then we give an analytical check by generalizing the method of moments \cite{viswanath1994recursion}.
Previously, a specific version of this method was applied to closed systems 
\cite{Parker:2018yvk, Bhattacharjee:2022ave}. We extract the analytical form of two sets of Lanczos coefficients, which correspond to the primary diagonal and primary off-diagonal elements of the Lindbladian matrix written in Krylov basis \cite{Bhattacharya:2022gbz}. We find that only the diagonal elements of the Lindbladian are sensitive to dissipation. We confirm our analytical results by semi-analytics in finite $q$ and large $N$ limit, and by directly implementing the numerical Arnoldi iteration in finite $q$ and finite $N$ limit. We emphasize that these three different approaches have their own advantages and disadvantages, in lieu of capturing the K-complexity behavior, for different types of systems. It is demonstrated that the large $q$ limit of the dissipative SYK model is a unique scenario where these approaches are equally viable.

The consistencies of all the results allow us to conjecture an operator growth hypothesis for the open systems, given by Eq.~\eqref{eq:anbn-krylov}. 
The analytical expressions of $a_n$ and $b_n$ are then used to derive analytical expressions for the Krylov basis wavefunctions, Krylov complexity, and Krylov variance. These expressions correspond to a special sub-class of the dissipative SYK model, where the dissipation strength is quadratically related to the random coupling $\mathcal{J}$. For weak dissipation, this model is a good approximation to the one considered in the rest of the manuscript. As a result, the Krylov complexity exhibits suppressed growth and saturation compared to the exponential growth for the closed systems. We identify the saturation timescale which grows logarithmically to the inverse of the dissipation strength. Although our results are based on the large $q$ analysis of the SYK model, we believe the claim of Eq.~\eqref{eq:anbn-krylov} to hold for any generic open quantum systems. In fact, some analysis on spin systems \cite{Bhattacharya:2022gbz, Liu:2022god} supports our claim. Based on the above observation, we argue the growth of K-complexity must upper bounds the growth of normalized out-of-time-ordered (OTOC), which might generalize the chaos bound \cite{Maldacena:2015waa} in open systems. Finally, we speculate a holographic interpretation of our results from the dual gravity side.

The manuscript is structured as follows.  In section \ref{sec:openkrylovreview} we review a few recently explored approaches to the K-complexity in open systems. In section \ref{setup}, we review the dissipative SYK model which will be our main playground. It was introduced in \cite{Kulkarni:2021xsx}, and more details can be found on \cite{Garcia-Garcia:2022adg, Kawabata:2022osw}. In section \ref{sec:largeq}, we derive the main analytical results in the large $q$ limit. Section \ref{sec:num} presents numerical evidence that these results still hold as an excellent approximation away from large $q$. In \ref{krylov} we derive the consequences in terms of the main results in terms of K-complexity growth and discuss implications on scrambling. We also give a heuristic explanation of our results from the gravitational side.
Finally, section \ref{concl} concludes the discussion with some open questions.

\section{Krylov complexity in open systems: general approaches}\label{sec:openkrylovreview}
The Krylov (K-)complexity of operators is mostly studied in closed Hamiltonian systems. Extending it to open systems is not straightforward at all, and so far there is no consensus on how one should proceed in general. Here, we restrict our attention to a restricted class of open dynamics that are described by a Lindblad master equation which admits the infinite temperature (maximally mixed) state $\rho_{\infty} \propto I$ as a stationary state and focus on the dynamics of operators in this ensemble. Even then, a few approaches are possible (and some have been explored), as we review in this section.

To set the common stage, let us recall the Lindblad equation in its general form \cite{Lindblad1976, Gorini}: 
\begin{align}
	\dot{\rho} = -i[H,\rho]+\sum_k \big[L_k \rho L_k^{\dagger}-\frac{1}{2}\{L_k^{\dagger} L_k,\rho\} \big]\,.
\end{align}
Here $H$ and $\rho$ are the Hamiltonian and the density matrix of the system, and $L_k$'s as the jump operators. As mentioned above, we shall assume that 
\begin{equation}
    \sum_k \big[L_k \rho_{\infty} L_k^{\dagger}-\frac{1}{2}\{L_k^{\dagger} L_k, \rho_{\infty}\} \big] = 0 \,,\, 
\end{equation}
where $\rho_{\infty} \propto I$ represents the infinite-temperature state. In this stationary state, it is reasonable to consider the operator dynamics, governed by \begin{align}
    \mathcal{O}(t) = e^{i\mathcal{L} t} \mathcal{O}\,,
\end{align}
where the Heisenberg-picture Lindbladian $\mathcal{L}$ is given by \cite{Bhattacharya:2022gbz}
\begin{align}
	\mathcal{L}  = \mathcal{L}_H + \mathcal{L}_D \,,\,~~~
	 \mathcal{L}_H \mathcal{O} = [H, \mathcal{O}] \,, ~~~
	  \mathcal{L}_D \mathcal{O} =  -i \sum_k \left[\mp L_k^{\dagger} \mathcal{O} L_k- \frac{1}{2}\{L_k^{\dagger} L_k, \mathcal{O}\} \right]\,. \label{negpos}
\end{align}
Here, we view $\mathcal{L}$ as a superoperator acting on the space of operators (the minus sign has to be taken when both $L_k$ and $\mathcal{O}$ are fermionic operators \cite{Liu:2022god}). It is written as a sum of a unitary Liouvillian $\mathcal{L}_H$ and a dissipation contribution $\mathcal{L}_D$. The space of operators is also endowed with the inner product 
\begin{equation}\label{eq:inner}
    \left( A | B \right) := \mathrm{Tr}[\rho_{\infty} A^\dagger B] = \mathrm{Tr}[ A^\dagger B] / \mathrm{Tr}[1]\,.
\end{equation}
Then, the autocorrelation function of an operator $\mathcal{O}$ can be defined similarly as in the closed-system context:
\begin{equation}\label{eq:Ctdef}
    \mathcal{C}(t) =  \left( \mathcal{O} | \mathcal{O}(t)  \right) = \left( \mathcal{O} | e^{i \mathcal{L} t}\vert \mathcal{O} \right)\,.
\end{equation}
Without loss of generality, we shall assume that $\mathcal{O}$ has unit norm: $ \left( A | B \right)  = 1 $.

\subsection{Using the closed Krylov basis}
Ref.~\cite{Li:2021kuv} proposed a simple way to extend the notion of K-complexity to the above open setting: we take the Krylov basis generated by the repeated action of the \textit{closed-system Liouvillian } $\mathcal{L}_H$. Recall that this is an orthonormal basis $\{ \mathcal{O}_0, \dots \mathcal{O}_n, \dots \}$ that spans the Krylov subspaces: $$ \mathrm{span}(\mathcal{O}_0,\dots, \mathcal{O}_n) = \mathrm{span}(\mathcal{O}, \mathcal{L}_H \mathcal{O}, \dots, \mathcal{L}_H^n \mathcal{O}) \,. $$ Assuming that the open-system operator dynamics remain in such a Krylov subspace, one can define its K-complexity in exactly the same way. Now, on this basis the full Lindbladian is represented as a general matrix: while $\mathcal{L}_H$ becomes tridiagonal, $\mathcal{L}_D$ enjoys no special property in general. Nevertheless, the authors of~\cite{Li:2021kuv} put forward conjectures on the matrix elements $(\mathcal{O}_n \vert \mathcal{L}_D \vert \mathcal{O}_m)$ motivated by numerical study, which we shall discuss below.

\subsection{Arnoldi iteration}
An alternative method, explored in Ref.~\cite{Bhattacharya:2022gbz}, considers another orthonormal basis $\{ \mathcal{V}_0, \dots \mathcal{V}_n, \dots \}$, which corresponds to the Krylov subspaces generated by the full open-system Lindbladian:
$$ \mathrm{span}(\mathcal{V}_0,\dots, \mathcal{V}_n) = \mathrm{span}(\mathcal{O}, \mathcal{L} \mathcal{O}, \dots, \mathcal{L}^n \mathcal{O}) \,. $$
This basis, as well as the representation of the Lindbladian in it, can be generated by the Arnoldi iteration that we recall now.

We start with an initial normalized vector $\mathcal{V}_0 \propto \mathcal{O}$. For $k=1, 2, \ldots$, we construct
\begin{align}
    |\mathcal{U}_k )= \mathcal{L} |\mathcal{V}_{k-1})\,.
\end{align}
Then, for $j=0$ to $n-1$, we perform the following iterations:
\begin{align}
    &1. ~h_{j,k-1} = ( \mathcal{V}_j|\mathcal{U}_k )\,.~~~~~~~~~~~~~~~~~~~~~~~~ ~~~~~~~~~~~~ ~~~~~~~~~~~~~~~~~~~~~~~~~   \nonumber \\
    &2. ~|\tilde{\mathcal{U}}_{k} ) =|\mathcal{U}_{k} )-\sum_{j=0}^{k-1} h_{j,k-1} |\mathcal{V}_{j} )\,. \nonumber \\
    &3. ~ h_{k,k-1} =  \sqrt{({\tilde{\mathcal{U}}_k|\tilde{\mathcal{U}}_k})}\,.
\end{align}
(Stop if $h_{k,k-1}=0$.) Otherwise define $\mathcal{V}_k$ as
\begin{align}
   | \mathcal{V}_k )= \frac{|\tilde{\mathcal{U}}_k )}{h_{k,k-1}}\,.
\end{align}
As a result, the Lindbladian is transformed into an upper Hessenberg form in the Krylov basis (sometimes called Arnoldi basis):
\begin{align}\label{Arnoldimatrix}
    \mathcal{L} \sim \begin{pmatrix}
h_{0,0} & h_{0,1} & h_{0,2} & \cdots & \cdots & h_{0,n}\\
h_{1,0} & h_{1,1} & h_{1,2} & \cdots & \cdots & h_{1,n}\\
0 & h_{2,1} & h_{2,2} & h_{2,3} & \cdots & \cdots\\
\cdots & 0 & h_{3,2} &\cdots & \cdots & \cdots\\
0 & \cdots & 0 &\cdots &\cdots & h_{n-1,n}\\
0 & 0 & \cdots &0 & h_{n,n-1} & h_{n,n}\\
\end{pmatrix}\,,
\end{align}
where $h_{m,n}= (\mathcal{V}_m|\mathcal{L}|\mathcal{V}_n)$. When $\mathcal{L}$ is a Hermitian superoperator, the Arnoldi iteration reduces to the Lanczos algorithm, and the above matrix becomes real symmetric and tridiagonal. However, for a general Lindbladian, these properties are not guaranteed. Thus, while the basis $\mathcal{V}_{k}$ allows to define a version of K-complexity, the Lindbladian is still relatively involved.

\subsection{The moment method} \label{momref}
A distinct approach is to generalize the ``moment method''~\cite{viswanath1994recursion}, which was previously studied in closed SYK \cite{Parker:2018yvk}, and its higher-order corrections \cite{Bhattacharjee:2022ave}. Rather than constructing a basis, this method focuses on the auto-correlation function, or its derivatives around $t = 0$, known as the moments:
\begin{align}
    \mathcal{C}(t)\, \stackrel{t>0} = \, \sum_{n=0}^{\infty} m_n \,\frac{(it)^n}{n!}\,. \label{autodef} 
\end{align}
Now, \textit{assuming} that the superoperator generating the dynamics is Hermitian, the Lanczos coefficients and the moments are related by a nonlinear transform. In terms of the Lanczos coefficients, the moments, encoded in the Green function, can be obtained as a continued fraction expansion~\cite{viswanath1994recursion, Parker:2018yvk}:
\begin{align}
  \sum_{n=0}^{\infty} z^n m_n  = \dfrac{1}{1 - a_0 z - \dfrac{b_1^2 z^2}{1- a_1 z - \dfrac{b_2^2 z^2}{1 - a_2 z - \dots }}} \,.
\end{align}
From the moments, the Lanczos coefficients can be calculated by a recursive algorithm~\cite{viswanath1994recursion}:
\begin{align}
M_{k}^{(0)} &= (-1)^k m_k \,,\, L_{k}^{(0)} = (-1)^{k+1} m_{k+1}\,, \nonumber \\ 
M_{k}^{(n)}  &= L_{k}^{(n-1)} - L_{n-1}^{(n-1)} \frac{M_{n-1}^{(n-1)}}{M_{k}^{(n-1)}}\,, \nonumber\\
L_{k}^{(n)}  &= \frac{M_{k+1}^{(n)}}{M_{n}^{(n)}} - \frac{M_{n-1}^{(n-1)}}{M_{k}^{(n-1)}}\,,\,~ k \ge n \,,
\nonumber \\
b_n &= \sqrt{M_{n}^{(n)}}\,, ~~~ a_n = - L_n^{(n)} \,. \label{mombn}
\end{align}
Note that the above relations are a bit more general than those used in \cite{Parker:2018yvk}, which are valid for the case where the odd moments vanish (this is the case for closed systems and a Hermitian operator $\mathcal{O}$, for example). 
Formally, we can apply the algorithm \eqref{mombn} to the moments of an open-system autocorrelation function, ignoring the above-mentioned assumption. Generically --- that is, barring the occurrence of division-by-zero errors --- we are guaranteed to obtain two sequences of complex numbers, $\{b_n\}_{n=1}^\infty, \{a_n\}_{n=0}^\infty$. However, it should be emphasized that they are \textit{not} the output of a Lanczos algorithm. In fact, the moment method generates a \textit{non-orthonormal} basis of the Krylov subspaces, in which $\mathcal{L}$ is represented as a symmetric tridiagonal matrix
\begin{align}\label{Arnoldimatrix1}
    \mathcal{L} \sim \begin{pmatrix}
a_0 & b_1 &  &  & \\
b_1 & a_1 & b_2 &  & \\
 & b_2 & a_2 & b_3 & \\
 &  & b_3 &\cdots & \cdots \\
 &  &  &\cdots & \cdots \\
\end{pmatrix}\,.
\end{align}
Thus, although we simplified the matrix representation of $\mathcal{L}$, it becomes problematic to define the notion of K-complexity using the non-orthonormal basis, which can be in practice quite singular (with respect to the inner product).

To conclude, we reviewed three different attempts to extend the K-complexity to an open-system context. None of them is entirely satisfactory; each of them solves certain problems by making others worse. There seems to be a list of ``wishes'' that cannot be simultaneously fulfilled in general. The main point of this paper is to show that all wishes can come true in \textit{certain} systems.

\section{The dissipative SYK} \label{setup}
In this section, we review and introduce a concrete system, which is a dissipative version of the Sachdev-Ye-Kitaev (SYK) model \cite{PhysRevLett.70.3339, Kittu}. The SYK model is a $(0+1)$-dimensional quantum mechanical model which consists of $N$ fermions where $q$ of them are interacting at a time. The SYK$_q$ Hamiltonian is given by \cite{PhysRevLett.70.3339}
\begin{align}
    H = i^{q/2} \sum_{i_1 < \cdots < i_q} J_{i_1 \cdots i_q} \psi_{i_1} \cdots \psi_{i_q}\,,
\end{align}
where $\psi_i$'s are Majorana fermions satisfying $\{\psi_a, \psi_b\} = \delta_{ab}$, and $J_{i_1 \cdots i_q}$ are random couplings, drawn from some Gaussian ensemble with zero mean $\braket{J_{i_1 \cdots i_q}} = 0$ and variance $\braket{J^2_{i_1 \cdots i_q}} = (q-1)! J^2/N^{q-1}$. The constant $J$ is a dimensionful parameter and sets the energy scale of the Hamiltonian. We will be particularly interested in the limit where we send $N\to\infty$, and then $q \to \infty$. In that limit, the model is quantum chaotic yet enjoys a high degree of analytical tractability; numerical studies are required more or less away from this limit. When $q$ is large, it is convenient to define a rescaled coupling constant:
\begin{equation}
    \mathcal{J}^2 = 2^{1-q} q J^2 \,.
\end{equation}
To introduce openness, we consider the following jump operators~\cite{Kulkarni:2021xsx}
\begin{align}
    L_{i} = \sqrt{\mu} \,\psi_{i}\,, ~~~~~~ i = 1, \cdots, N\,. \label{jumpop}
\end{align}
where $\mu \geq 0$ is the dissipation strength or that of the system-environment coupling. The dissipation operators are fermionic and Hermitian. The Hermiticity ensures that the infinite-temperature density matrix $\rho_{\infty} \propto I$ \cite{Kulkarni:2021xsx, Sa:2021tdr, Garcia-Garcia:2022adg} is a stationary state. It will be useful to note that, the action of the dissipative part of the Lindbladian $\mathcal{L}_D$ is very simple on ``Majorana string'' operators. For any $s > 0$ and $ 1 \le i_1 < \dots < i_s \le N$, we have
\begin{equation}
    \mathcal{L}_D (\psi_{i_1} \dots \psi_{i_s}) = i \mu  s \, (\psi_{i_1} \dots \psi_{i_s}) \label{eq:LDMajorana} \,.
\end{equation}
See Appendix \ref{appa} for the derivation. In other words, the effect of the dissipation is to annihilate a Majorana string with a rate proportional to its size $s$ and to $\mu$ (recall that the operator evolution is generated by $i \mathcal{L}$). Therefore, $\mathcal{L}_D$ and $\mathcal{L}_H$ have competing effects as the latter makes the operator size grow. In the large $q$ limit, we rescale $\mu$ as well, so that
\begin{equation}
    \tilde\mu = \mu q \,, \label{mudef}
\end{equation}
is of order unity~\cite{Kulkarni:2021xsx}.

Often, the dynamics of the operators (as well as of the density matrix) is seen as taking place in the double Hilbert space $\mathcal{H} \otimes \mathcal{H}^{*} = \mathcal{H}_{L} \otimes \mathcal{H}_{R}$ through the lens of Choi-Jamiolkowski (CJ) isomorphism \cite{CHOI1975285, JAMIOLKOWSKI1972275}. In this regard, the operator dynamics is generated by the non-Hermitian Hamiltonian~\cite{Garcia-Garcia:2022adg}:
\begin{align}
    H = i H_{L}^{\mathrm{SYK}} - i (-1)^{q/2}\, H_{R}^{\mathrm{SYK}} 
    + i \mu \sum_{j} \psi_{L}^j \psi_{R}^j\,. \label{haam}
\end{align}
This can be understood as two SYK (with relative phase between them) interacting with the coupling $H_{\mathrm{int}} = i \mu \sum_{j} \psi_{L}^j \psi_{R}^j$. The state $\ket{\mathcal{I}}$, which is the image of $\rho_{\infty} = I$ under the CJ-isomorphism, is the ground state of $H_{\mathrm{int}}$ and well as $H$. Note that this is slightly different than Ref.~\cite{Maldacena:2018lmt}, where the Hamiltonian is Hermitian (the imaginary factor is absent in the first two terms). Hence, the dissipative dynamics are equivalent to non-Hermitian dynamics  governed by the above Hamiltonian.

\section{Large $q$ exact result}\label{sec:largeq}
We are in a position to state the main result of this work: for the dissipative SYK in the large $q$ limit with $\mathcal{O}\propto \psi_1$, the three approaches in Section~\ref{sec:openkrylovreview} \textit{exactly coincide}, in the sense that:
\begin{enumerate}
    \item The closed-system Krylov basis $\{\mathcal{O}_n\}$ is identical to the open-system one $\{\mathcal{V}_n\}$.
    \item The Arnoldi iteration yields a symmetric tridiagonal matrix, whose matrix elements
    are equal to the Lanczos coefficients obtained by the moment method. In particular, we have
    \begin{align}
      h_{n,n-1} = h_{n-1,n} = b_n \,,\,~~  h_{n,n} = a_n\,.
    \end{align}
\end{enumerate}
Furthermore, these coefficients are exactly calculated:
\begin{align}
&a_n = i \tilde{\mu} n + O(1/q) \,,\,~~ \tilde{\mu} := \mu q \,,\, \label{eq:an-main} \\ &b_n =  \begin{cases} \mathcal{J}\sqrt{2/q}  \,      & \,n = 1\,,\\
    \mathcal{J}\sqrt{n(n-1)} + O(1/q) \,   & \,n > 1\,. \label{eq:bn-main}
  \end{cases} 
 \end{align}
Here, the results on $b_n$ are identical to the closed-system one obtained in \cite{Parker:2018yvk}, which is expected from 1 above. The result on $a_n$ is new and shows that the diagonal elements are imaginary, and grow linearly with $n$ with a slope set by the dissipation strength~\cite{Liu:2022god}. 
 
In the rest of this section, we first present a direct derivation of the above results. Then, a nontrivial check of the Lanczos coefficients formulae \eqref{eq:an-main} and \eqref{eq:bn-main} via the moment method will be exposed. Finally, numerical results away from the ideal limit will be discussed.

\subsection{Operator size concentration}\label{sec:sizeargument}
It turns out that the above statements are all corollaries of the following  ``operator size concentration'' property of the Krylov basis elements of the \textit{closed} SYK model in the large $q$ limit: the $n$-th basis operator $\mathcal{O}_n$ is a linear combination of Majorana strings of the same size 
\begin{equation}\label{eq:OSC}
    \mathcal{O}_n = \sum_{i_1< \dots < i_s} c_{i_1, \dots, i_s} \psi_{i_1} \dots \psi_{i_{s}} + O(1/q) \,,\, 
\end{equation}
where 
\begin{equation}
    s = n (q-2) + 1 \,.
\end{equation}
Here $s$ is the size and $n$ is the generation, according to the nomenclature of Ref.~\cite{Roberts:2018mnp}. Indeed, recall \eqref{eq:LDMajorana} that Majorana strings of size $s$ are degenerate eigen-operators of the dissipative part of Lindbladian, with eigenvalue $i \mu s$. Thus, \eqref{eq:OSC} implies 
\begin{equation}
    \mathcal{L}_D \, \mathcal{O}_n  = i \mu s \, \mathcal{O}_n = i \tilde{\mu} n\, \mathcal{O}_n + O(1/q) \,,
\end{equation} 
where $\tilde{\mu}$ is defined through \eqref{mudef}. This, combined with the known action of $\mathcal{L}_H$, implies the statements above as well as \eqref{eq:an-main} and \eqref{eq:bn-main} immediately. 

It remains to prove operator size concentration~\eqref{eq:OSC}, which concerns the closed SYK model in the large $q$ limit. It is known~\cite{Roberts:2018mnp, Parker:2018yvk} that one can study the operator dynamics and implement the Lanczos algorithm directly in the large $N$ limit using ``open'' melon diagrams. Our proof will rely on this combinatorial approach, which we briefly review. For simplicity we set $\mathcal{J} = 1/\sqrt{2}$ below.

Let us iteratively apply $\mathcal{L}_H$ to $\psi_1$ and describe graphically the operators generated. It will be convenient to decompose the Liouvillian as  \cite{Caputa:2021sib}
\begin{equation}\label{eq:Lpm}
    \mathcal{L}_H = \mathcal{L}_+ + \mathcal{L}_-\,,
\end{equation}
where $\mathcal{L}_\pm$ are the operator-size increasing (decreasing, respectively) contribution to $\mathcal{L}_H$. In other words,  $\mathcal{L}_+$ is defined such that for two Majorana strings of sizes $s$ and $t$,
\begin{equation}
  ( \psi_{i_1} \dots \psi_{i_t} \vert \mathcal{L}_+ \vert \psi_{i_1} \dots \psi_{i_s} )
  = \begin{cases}  ( \psi_{i_1} \dots \psi_{i_t} \vert   \mathcal{L}_H \vert \psi_{i_1} \dots \psi_{i_s} ) & t > s  \\
  0 & \text{otherwise}.
  \end{cases}
\end{equation}
 $\mathcal{L}_-$ is then defined as the Hermitian conjugate of $ \mathcal{L}_+$. By applying $\mathcal{L}_+$ instead of $\mathcal{L}_H$, we can focus on the new operators generated at each step. 

At the first step,
\begin{equation}
    \mathcal{L}_+ \psi_1  \propto  \includegraphics[scale=.5,valign=c]{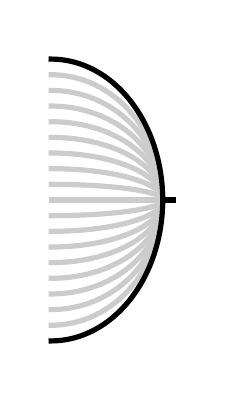} = \includegraphics[scale=.5,valign=c]{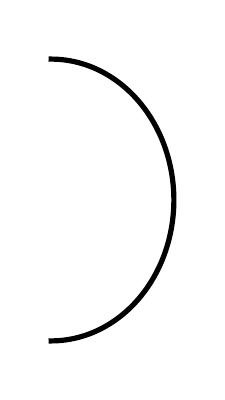} \,.
\end{equation} 
Here, we first represent $  \mathcal{L}_H \psi_1$, which is a $(q-1)$-body operator generated from the $\psi_1$, by the half of a melon diagram as in \cite{Roberts:2018mnp}. Since $q$ is large, we omit all (grey) propagators but two representatives, and also the tip of the melon. Hence the diagram simplifies to a simple arc.

With this in mind, the next few operators generated will be the following: 
\begin{align}
   &  \mathcal{L}_+^2 \psi_1 \propto \includegraphics[scale=.5,valign=c]{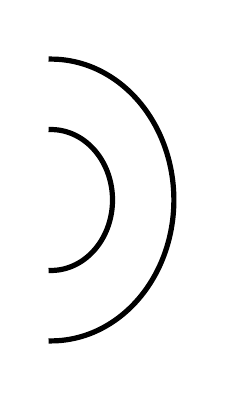}  \,, \label{eq:LH2}\\
    &  \mathcal{L}_+^3 \psi_1 =  c_3\includegraphics[scale=.5,valign=c]{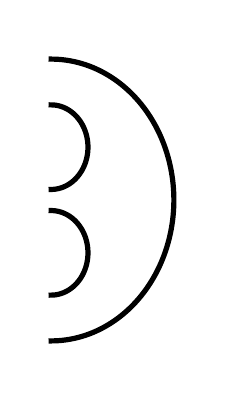} + c_4 \includegraphics[scale=.5,valign=c]{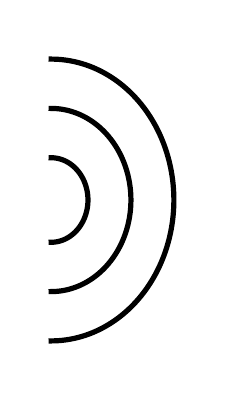}  \,, \label{eq:LH3}\\
    & \mathcal{L}_+^4 \psi_1 =  c_5\includegraphics[scale=.5,valign=c]{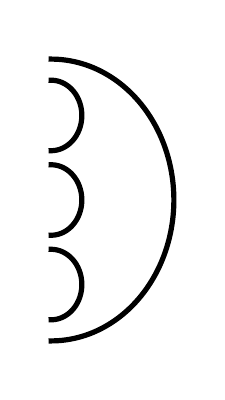} + c_6 \includegraphics[scale=.5,valign=c]{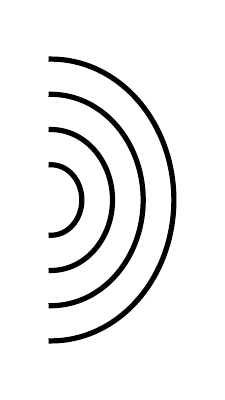} + c_7 
    \includegraphics[scale=.5,valign=c]{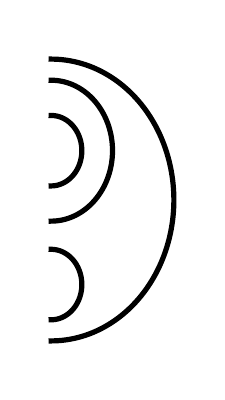} + c_8
     \includegraphics[scale=.5,valign=c]{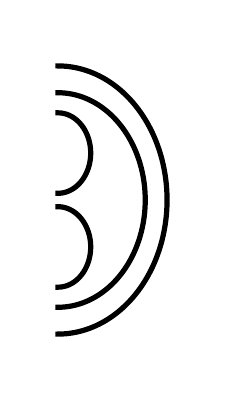} ,\label{eq:LH4}
\end{align}
and so on. Every time, the $\mathcal{L}_+$ acts on one of the Majorana's inside an arc, and transforms it into $(q-1)$ Majorana's represented by a new arc, which we view as a ``child'' of the former. We can equivalently view the operators appearing in $ \mathcal{L}_+^n \psi_1$ as rooted and unmarked trees with $n$ vertices (a vertex = an arc). We remark that equations \eqref{eq:LH2} and \eqref{eq:LH4} are operator identities that are valid before averaging over the disordered couplings $J_{i_1 \cdots i_q}$, \textit{modulo} terms that will have negligible contributions to any disordered-averaged observables in the large $N$ limit. Each arc involves a dangling disorder line. The subleading terms are neglected by considering only open melon diagrams. The observables can be constructed by closing the open melon diagrams, that is, taking the inner product with another operator and averaging over disorder.

So far, we have been elusive about the prefactors. A neat way to keep track of them is to order the arcs in the diagrams, such that a vertex comes always before its children. For example, the diagram with prefactor $c_7$ above can be ordered in three ways:
$$
    c_7 
    \includegraphics[scale=.5,valign=c]{sykq-op7.pdf} =  \includegraphics[scale=.6,valign=c]{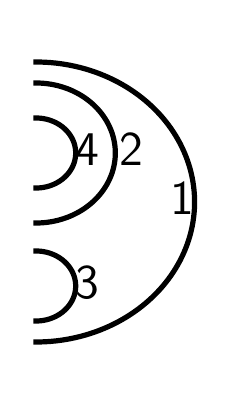} +
    \includegraphics[scale=.6,valign=c]{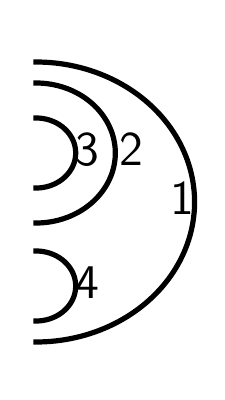} +
    \includegraphics[scale=.6,valign=c]{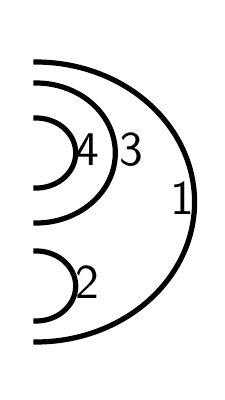} \,.
$$
In other words, the amplitude of an unmarked diagram in $\mathcal{L}_+^n \psi_1$ is given by the ways it can be built by adding one arc at a time. Note that the terms of the right-hand side are the same operator; the different orderings enumerate its amplitude (multiplicity) in $\mathcal{L}_+^n \psi_1 $. As a consequence, 
\begin{equation}
    \mathcal{L}_+^n \psi_1 = \sum \left[ \text{ordered diagrams of $n$ vertices} \right] \,.  \label{eq:Lnismarkedtreee}
\end{equation}
We now come to the essential claim of this section. For any $n \ge 1$, we have 
\begin{equation}
    \mathcal{L}_- \mathcal{L}_+^{n+1} \psi_1 = \frac12 n (n + 1) \mathcal{L}_+^{n} \psi_1 \,. \label{eq:central}
\end{equation} 
This identity can be proved diagrammatically. By \eqref{eq:Lnismarkedtreee} and the definition of $\mathcal{L}_-$, the left-hand side is a sum of ordered diagrams of $n+1$ arcs with one \textit{childless} arc marked. For example, 
\begin{align}
    \mathcal{L}_- \mathcal{L}_+^{3+1} \psi_1 = \,\mathcal{L}_-  \includegraphics[scale=.6,valign=c]{sykq-op7m2.pdf} + \dots   = \includegraphics[scale=.6,valign=c]{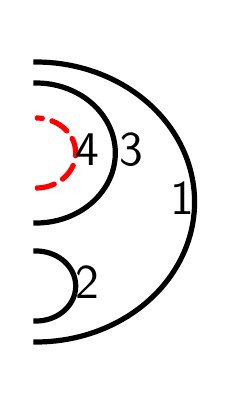} + \includegraphics[scale=.6,valign=c]{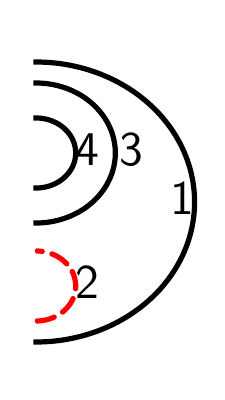} + \dots \,.
\end{align}
In the second line, the marked childless arc is represented by a red dashed line. One should view it as being removed by the action of $\mathcal{L}_-$. 

Now, to prove \eqref{eq:central}, we shall show that the removal of the marked arc gives rise to an $n(n+1)/2$-to-one correspondence between the ensemble of marked and ordered diagrams with $n+1$ arcs, and that of ordered diagrams with $n$ arcs. To do this, consider a marked and ordered diagram of $3+1$ arcs, say, $\includegraphics[scale=.6,valign=c]{sykq-op7m2c1.pdf}$. Removing the marked arc gives rise to the following ordered diagram with $3$ arcs: $$ \includegraphics[scale=.6,valign=c]{sykq-op7m2c1.pdf} \mapsto \includegraphics[scale=.6,valign=c]{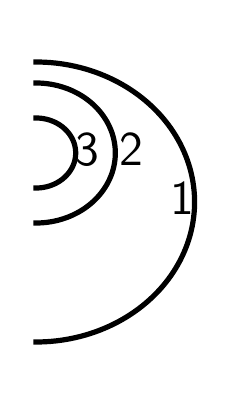}  \,. $$ This removal map is not one-to-one, since we lost information about the marked arc, namely, its parent's index $p \in \{ 1, \dots, n\}$, and its order $q \in \{p + 1, \dots, n+1 \}$ (since it must come later than its parent). In the above example, $(p,q) = (1,2)$ in the example. It is not hard to see that the datum of 
\begin{equation}\label{eq:tupledef}
     (p,q) \text{ s.t. } p \in \{1, \dots, n\} \,,\, q \in \{p+1, \dots, n+1\} \,,
\end{equation} allows to reconstruct uniquely the marked ordered diagram with $n+1$ arcs from any ordered diagram of $n$ arcs. For example,
$$ \left[ \includegraphics[scale=.6,valign=c]{sykq-op7m3.pdf}, p=2, q=3 \right]  \mapsto   \includegraphics[scale=.6,valign=c]{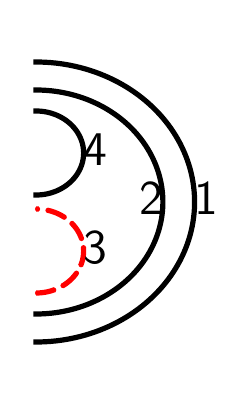} \,. $$
Observe that removing the marked arc on the right-hand side gives back the diagram on the left-hand side. 
Since there are 
$$ 1 + 2 + \dots + n =  \frac12 n (n+1)$$
such tuples satisfying \eqref{eq:tupledef}, the removal map is a $n (n+1)/2$-to-one correspondence. This concludes the combinatorial proof of \eqref{eq:central}.

The key consequence of \eqref{eq:central} is that, the Krylov basis generated by $\mathcal{L}_H$ and $\psi_1$ is essentially $\{\mathcal{L}_+^n \psi_1 \}$, up to normalization: 
\begin{equation}\label{eq:OnisLplus}
     \mathcal{O}_n \propto \mathcal{L}_+^n \psi_1 \,.
\end{equation}
To see this, we first notice that $\{\mathcal{L}_+^n \psi_1 \}$ is an orthogonal set, since $\mathcal{L}_+^n \psi_1$ is a linear combination of Majorana strings of length $ s = n(q-2) + 1$, and Majorana strings of distinct lengths are orthogonal under the inner product~\eqref{eq:inner}. Moreover, by \eqref{eq:Lpm} and \eqref{eq:central}, for $n > 1$, 
\begin{align}\label{eq:LhLp}
    \mathcal{L}_H \mathcal{L}_+^n \psi_1 = 
   (\mathcal{L}_+ + \mathcal{L}_-) \mathcal{L}_+^n  \psi_1
   =  \mathcal{L}_+^{n+1} \psi_1 + \frac12 n (n-1) \mathcal{L}_+^{n-1} \psi_1 \,.
\end{align}
The edge cases $n=0,1$ can be taken care of by explicit calculation:
\begin{equation}
   \mathcal{L}_H \psi_1 = \mathcal{L}_+ \psi_1 \,,\,~~ \mathcal{L}_H \mathcal{L}_+ \psi_1  = \mathcal{L}_+^{2} \psi_1 + \frac1q  \psi_1\,.
\end{equation}   
Therefore, the Liouvillian $ \mathcal{L}_H  $ is represented as a tridiagonal matrix in the basis $\{\mathcal{L}_+^n \psi_1 \}$. Combined with its orthogonality, we conclude that it must be identical to the Krylov basis up to normalization, \eqref{eq:OnisLplus}. In fact, \eqref{eq:LhLp} implies that the Lanczos coefficients $b_n = \sqrt{n (n-1) / 2}$ for $n> 1$, as announced in \eqref{eq:bn-main} (recall that we set $\mathcal{J}=1/\sqrt{2}$).\footnote{The square-root comes from the appropriate normalization of $\mathcal{L}_+^n \psi_1 $, to get $\mathcal{O}_n$. The way to proceed with this is to consider the norm, $\Vert \mathcal{L}_+^n \psi_1 \Vert^2 =(\psi_1 \vert \mathcal{L}_-^n \mathcal{L}_+^n \vert \psi_1) = \frac{n (n-1)}2 \frac{(n-1) (n-2)}2  \dots (\psi_1 \vert \psi_1)$, which gives the normalization squared.} Thus, we have provided an independent, diagrammatic derivation of this nontrivial result (the derivation in \cite{Parker:2018yvk} relied on the moment method and knowledge about tangent numbers, see also below).

Eq.~\eqref{eq:OnisLplus} implies operator size concentration~\eqref{eq:OSC} immediately: since $\mathcal{L}_+^n \psi_1$ is by construction a linear combination of Majorana strings of length $s= n(q-2) + 1$, the same is true for the $n$-th Krylov basis element of the large $q$ SYK model. This concludes the derivation of the main results announced at the beginning of the section. 

\subsection{Consequence on moments} \label{lanczos}
A consequence of the main results is that the Lanczos coefficients \eqref{eq:an-main} and \eqref{eq:bn-main} are related to the autocorrelation function of the dissipative SYK in the large $q$ limit. The latter has been computed explicitly by solving a Schwinger-Dyson equation~\cite{Kulkarni:2021xsx}:\footnote{
We remark that the large $q$ expansion above is valid for $t \ll t_c$, where $t_c$ is such that $|g(t_c)| = O(q)$, or $t_c \sim q$, for large $q$. 
Beyond that time scale, the expansion \eqref{eq:Ct} breaks down, and one would need a different approximation \cite{Kawabata:2022osw}.}
\begin{align}
  &  \mathcal{C}(t) =  1 + \frac1q g(t) + O(1/q^2) \,, \label{eq:Ct} \\
  & g(t) =  \log \left[\frac{\alpha^2}{\mathcal{J}^2\cosh^2(\alpha t + \gamma)}\right] \,,\, t > 0 \,,
\end{align}
where we recall that $\mathcal{J}^2 = 2^{1-q} q J^2, \tilde{\mu} = \mu q$, and the other variables are given by 
\begin{align}
     \alpha = \sqrt{\bigg( \frac{\tilde{\mu}}{ 2 }\bigg)^2 +  \mathcal{J}^2}\,, ~~ \gamma = \arcsinh \bigg(\frac{\tilde{\mu}}{ 2 \mathcal{J}}\bigg)\,. \label{pardef}
\end{align}
In the closed-system case $\mu = 0$, we recover the known result $g(t) = 2 \ln \sech (\mathcal{J} t)$~\cite{Maldacena:2016hyu}, and the associated moments are essentially the tangent numbers, and the Green function is known to have a continued fraction expansion given by the $b_n$'s~\cite{Parker:2018yvk}. When $\mu > 0$, the moments are considerably more involved. Setting $\mathcal{J} = 1$ without loss of generality, the moments have the large $q$ expansion
\begin{align}
    m_n = \frac{2}{q} \tilde{m}_n + O(1/q^2) \,,\,~~ n \ge 1 \,,
\end{align}
and $\tilde{m}_n$ is a polynomial of $u := i \tilde\mu$. For example, $\tilde m_1 = u/2$,
\begin{align}
      \tilde m_2 &= 1\,, \\
      \tilde m_3 &= u \,,\\
      \tilde m_4 &= u^2 + 2\,, \\
      \tilde m_5 &= u^3 + 8 u \,,\\
      \tilde m_6 & = u^4 + 22 u^2 + 16 \,,\\
       \tilde m_7 & = u^5 + 52 u^2 + 136 \,,\\
        \tilde m_{8} & = u^6+114 u^4+720 u^2+272 \,,
\end{align}
and so on. The coefficients appear to coincide with the so-called triangle ``$T(n,k)$'', see \cite{A101280} and references therein for further mathematical facts about it. Our main result then implies the following continued fraction formula 
 \begin{align} \label{eq:cfformula}
     \sum_{n \ge 0} \tilde{m}_{n+2} z^{n} =
      \dfrac{1}{1 - u z - 
     \dfrac{1 \times 2 z^2}{1- 2 u z - \dfrac{2 \times 3 z^2}{1 - 3 u z -   \dfrac{3 \times 4 z^2}{1 - 4 u z - \dots} }}}   \,.
 \end{align}
 One may check this formula order by order in $z$. We are not aware of an independent ``elementary'' proof. See however discussion around Eq.~\eqref{eq:sykqexact} below. 

\section{Finite $q$ numerics}\label{sec:num}
In this section, we turn to examine how the simple ``ideal'' scenario established above at the $q\to\infty$ limit is affected at finite $q$. 

\subsection{Large $N$: operator size distribution}\label{sec:largNOSC}
Our main results at the large $q$ limit rely on the operator size concentration property of the closed-system Krylov basis. A natural way to anticipate how these results survive at finite $q$ is to examine the operator size distribution \cite{Roberts:2018mnp, Qi:2018bje}. This can still be done in the large $N$ limit, by extending the diagrammatic approach to large $N$ operators~\cite{Roberts:2018mnp, Parker:2018yvk} described in the previous section to an arbitrary value of $q$: a finite $q$ imposes that every arc can have at most $q-1$ (direct) children. Coding this in a computer, we may generate all the large $N$ operators up to a certain size, which allows us to implement the Lanczos algorithm in the large $N$ limit and construct explicitly the Krylov basis operators, up to a certain value of $n$. (As the number of operators increases exponentially in $n$, we are limited to $n \approx 20$; fortunately, this is enough to observe the asymptotic behaviors.)

We then measure the operator size distribution of the Krylov basis operators $\mathcal{O}_n$. In a finite size system, this is defined as follows~\cite{Roberts:2018mnp}: writing $\mathcal{O}_n$ as a sum of Majorana strings
\begin{equation}
    \mathcal{O}_n = \sum_s \sum_{i_1< \dots < i_s} c_{i_1, \dots, i_s} \psi_{i_1} \dots  \psi_{i_s} \,,
\end{equation}
the size distribution is then given by
\begin{equation}
    P_s = \sum_{i_1< \dots < i_s} |c_{i_1, \dots, i_s}|^2 \,.
\end{equation}
In the large $N$ representation, an operator is a sum of diagrams, each representing a sum of Majorana strings with definite size $s$, related to the number of arcs $n$ as $s = (q-2) n + 1$. The weights $|c_{i_1, \dots, i_s}|^2$ depends on the random coupling coefficients $J_{i_1 \cdots i_q}$. Yet, one can show that each diagram after disorder averaging and summing over fermion flavors, each diagram gives the same weight. Hence, the size distribution can be readily measured at large $N$. Using the size distribution, we can obtain readily the average operator size $\braket{s} =: \sum_s s P_s $ as well as higher moments.

\begin{figure}[t]
\subfigure[]{\includegraphics[height=4.6cm,width=0.47\linewidth]{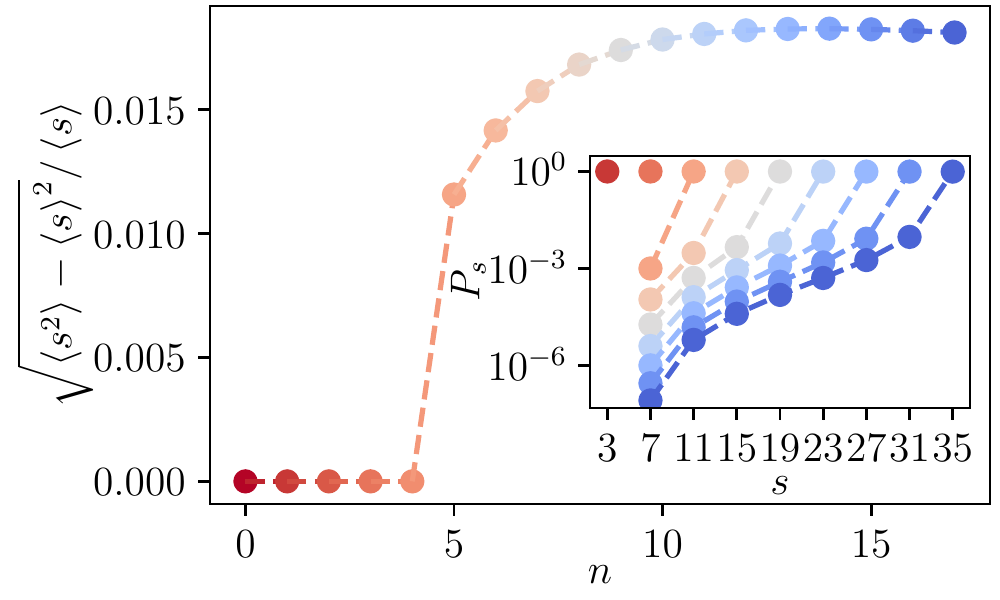}}
\hfill
\subfigure[]{\includegraphics[height=4.8cm,width=0.47\linewidth]{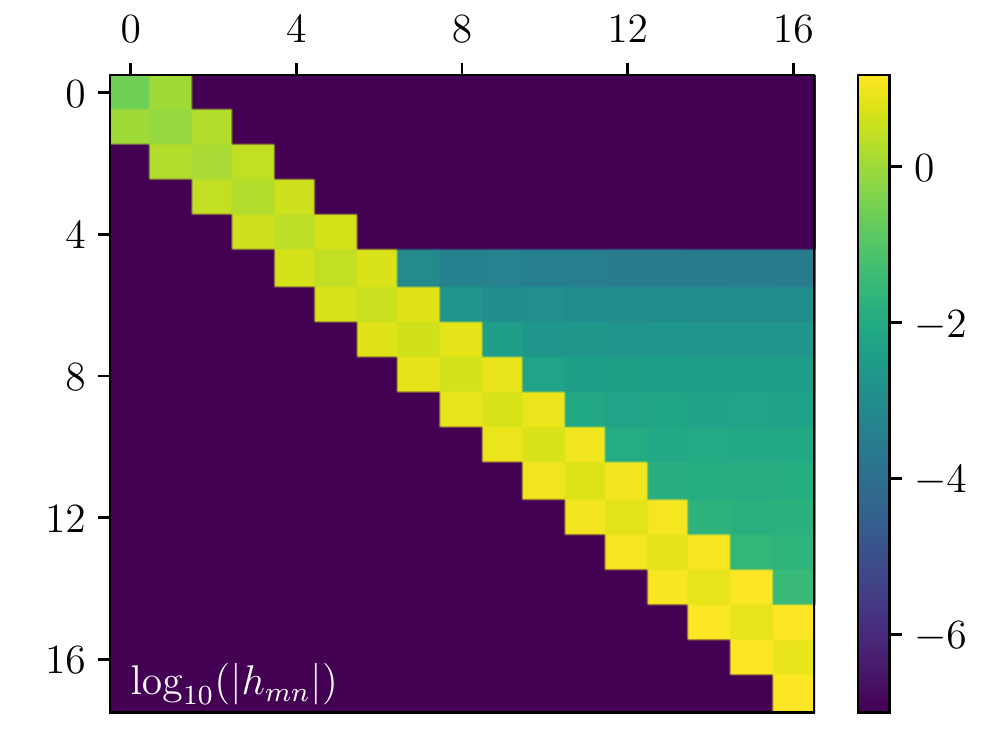}}
\caption{(a) Operator size distribution of the Krylov basis operators $\mathcal{O}_n$ in large $N$, $q=4$ SYK model. In the main plot, we plot the standard deviation of the operator size divided by the average size of $\mathcal{O}_n$ as a function of $n$. The inset shows the whole operator size distribution for a few Krylov basis operators (the value of $n$ is given by the color code which is the same as in the main). The dashed lines are a guide to the eye. The distribution is strongly peaked at the largest operator size $s = (q-2) n + 1$ present in $\mathcal{O}_n$, so that the standard deviation is $\le 0.016$ times the mean for all $n \le 17$, and appears to tend to a constant as $n$ increases.  (b) Magnitude of the matrix elements of the matrix $h_{m,n}$ resulting from the Arnoldi iteration in the large $N$, $q=4$ dissipative SYK model with $\tilde{\mu} = 1$ and $J = 1$. The presence of the secondary off-diagonal elements in the upper Hessenberg form indicates the finite $q$ effect.} \label{fig:opsize_syk4}
\end{figure}

In Fig.~\ref{fig:opsize_syk4} (a), we plot the results for $q = 4$. We observe that the operator size concentration property holds exactly if and only if $n = 0,\dots, 4$: only the first five Krylov basis operators are a sum of Majorana's of a definite size $s = (q-2) n +1$. For larger $n$, the \textit{exact} operator size concentration is indeed only a property of the large $q$ limit.\footnote{We have checked a few other values of $q$ and observed that the property always breaks down at $n=5$.} Nevertheless, the distribution of operator size is highly peaked at $s = (q-2)n+1$, which is the maximal operator size present in $\mathcal{O}_n$. The weight of operator sizes $s' < s$ decays exponentially fast in $s - s'$, see the inset of Fig.~\ref{fig:opsize_syk4} (a). As a result, the mean operator size 
$\left<s\right> \approx (q-2)n+1$ grows linearly in $n$. Its standard deviation also grows linearly with $n$, yet much more slowly than the mean value. Our numerics indicates the standard deviation-mean ratio tends to be a small constant:
\begin{equation}
  \frac{\sqrt{ \left< s^2 \right> -  \left< s \right>^2 }}{\left< s \right>} \to C \approx 0.015 \,,
\end{equation}
for $q=4$ SYK. Data for other values of $q$ suggest that the constant decays as $C \approx 0.06 / q$.

In summary, we provided significant evidence indicating that operator size concentration is an excellent approximation in the SYK model for any values of $q$. We may further rationalize this optimism by observing that the operator size concentration holds exactly for $q = 2$ (since only Majorana strings of length $1$ are generated) as well as for $q = \infty$; hence, it cannot be too wrong at intermediate values of $q$.

\subsection{Large $N$: Arnoldi iteration}\label{sec:largeNarnoldi}
As a consequence of operator size concentration, we expect that the Arnoldi iteration should result in a matrix $h_{m,n}$ close to the ``ideal scenario''~\cite{Bhattacharya:2022gbz,Liu:2022god}. That is, to a good approximation, 
\begin{enumerate}
    \item $[h_{mn}]$ should be a symmetric tridiagonal matrix,
    \item its diagonal elements are imaginary and grow linearly in $n$ with a pre-factor proportional to $\mu$:
    \begin{equation}
        h_{n,n} \simeq i \mu \chi n \,. \label{eq:hnn-ideal}
    \end{equation}
    Here $\chi$ is some $\mu$-independent proportionality constant.
    \item its primary off-diagonal elements grow linearly in $n$:
    \begin{equation}
        h_{n-1,n}\simeq \alpha n \,,\label{eq:hn1n-ideal}
    \end{equation}
    where $\alpha$ is independent of $\mu$.
\end{enumerate}
This is indeed what we found from our large $N$ numerics. In Fig.~\ref{fig:opsize_syk4}(b) we plot the magnitude of the matrix elements $h_{m,n}$ obtained from the Arnoldi iteration. As expected, the largest matrix elements are on the diagonal and primary off-diagonals; the other nonzero elements, that is, $h_{m,n}$ for $m < n -1$, are about two orders of magnitudes smaller (both the dissipation rate and the unitary coupling constant is of order one, $\tilde\mu = J = 1$).

\begin{figure}[t]
\subfigure[]{\includegraphics[height=4.6cm,width=0.47\linewidth]{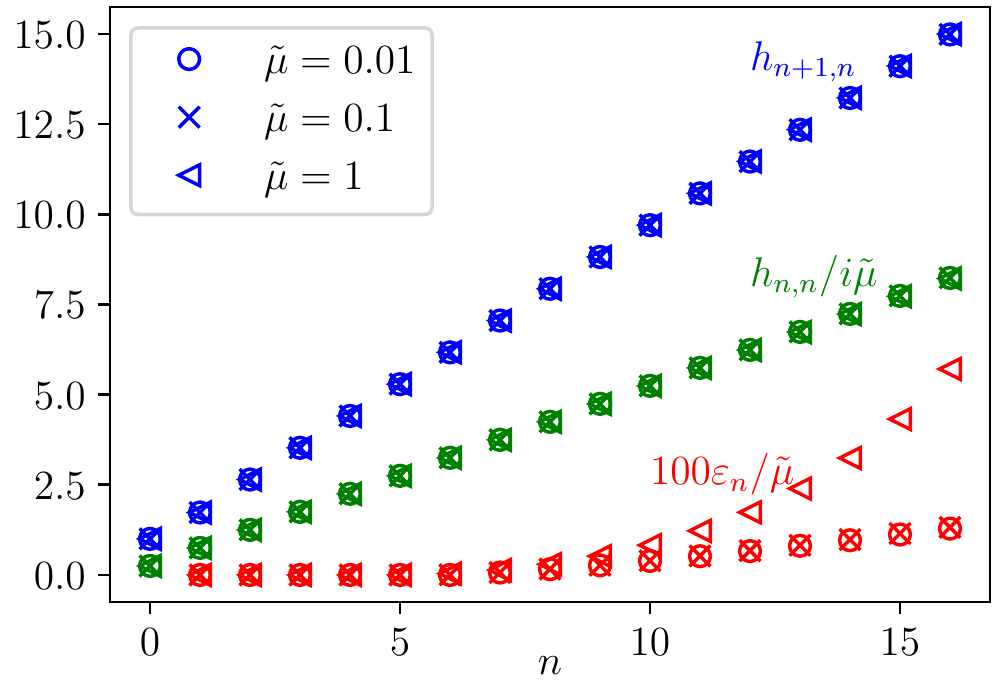}}
\hfill
\subfigure[]{\includegraphics[height=4.8cm,width=0.47\linewidth]{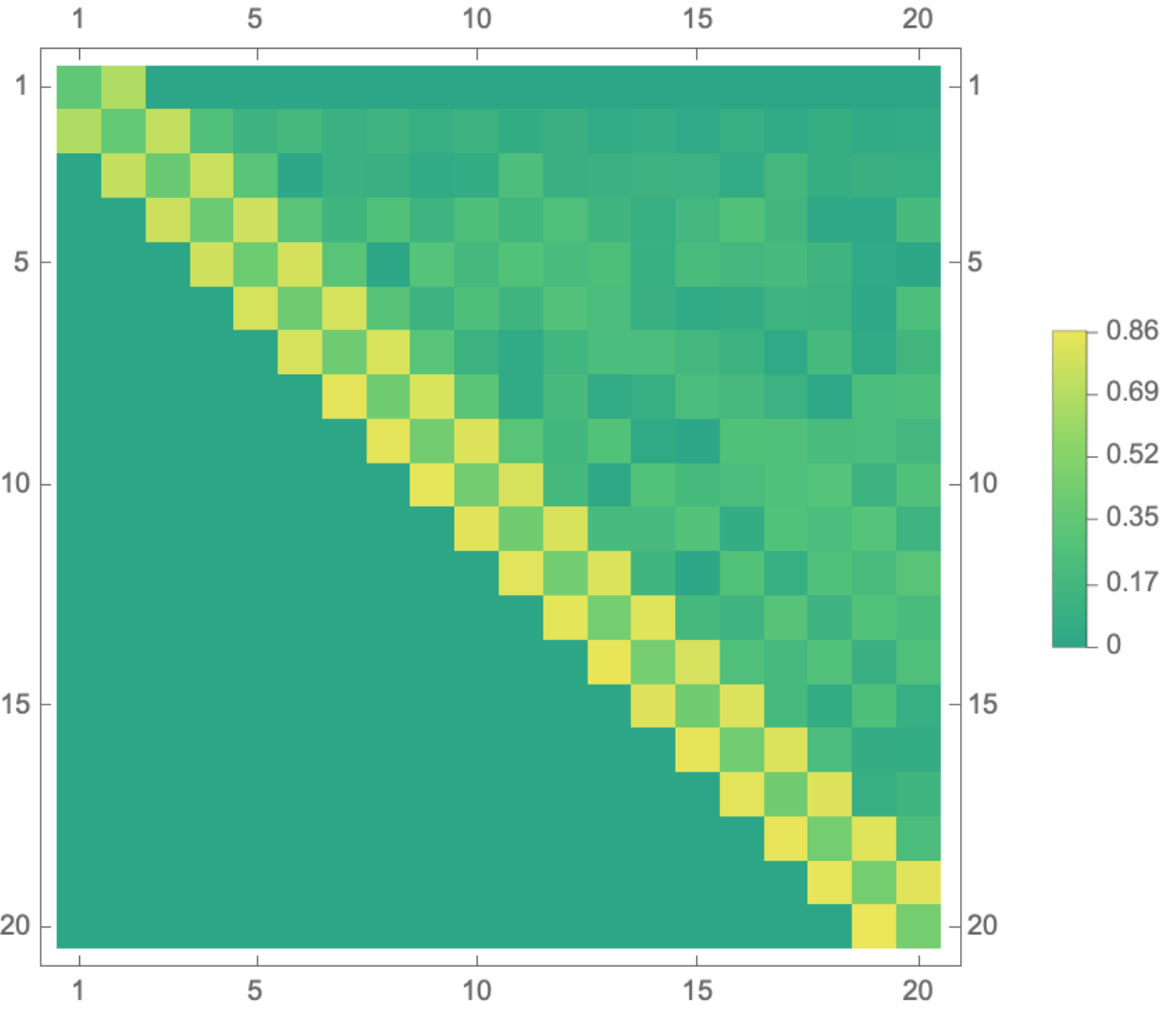}}
\caption{(a) The diagonal and primary off-diagonal elements in the matrix $h_{m,n}$ resulting from the Arnoldi iteration in the large $N$, $q=4$ dissipative SYK model with $\Tilde{\mu} = 1, 0.1, 0.01$ and $J = 1$. The markers indicate the value of $\tilde\mu$. We also plot the quantity $\varepsilon_n$ defined in \eqref{eq:errdef} which measures the distance to the ideal scenario (it is multiplied by $100$ for visibility). The diagonal elements and $\varepsilon_n$ are rescaled by $\tilde{\mu}$ to display the collapse. (b) $\mathrm{SYK}_4$ (i.e., $q=4$) Lindbladian matrix (single realization) in Krylov (Arnoldi) basis for dissipation $\mu = 0.01$. We fix $J = 1$ and the system size $N=18$. The presence of the secondary off-diagonal elements in the upper Hessenberg form is due to both $q$ and $N$ being finite, and to the absence of disorder averaging; compare with Fig.~\ref{fig:opsize_syk4} (b).} \label{fig:arnoldi-syk4-anbn}
\end{figure}

Next, let us examine more quantitatively how close the matrix is to the ``ideal scenario''. In Fig.~\ref{fig:arnoldi-syk4-anbn} (a), we plot the primary off-diagonal elements $h_{n+1,n}$ (which are real), the imaginary part of the diagonal elements $\mathrm{Im}(h_{n,n})$ (they turn out purely imaginary), and also an ``error'' defined as
\begin{equation} \label{eq:errdef}
    \varepsilon_n^2 := |h_{n-1,n} - h_{n,n-1}|^2 + \sum_{k < n-1} |h_{k,n}|^2 \,,~~ n > 0 \,.
\end{equation} 
By definition, the error vanishes if and only if $h$ is symmetric. We observe almost perfect linear growth in $n$ for both $h_{n+1,n}$ and $\mathrm{Im} (h_{n,n})$. The growth rate of $\mathrm{Im}(h_{n,n})$ is proportional to the dissipation rate $\tilde{\mu}$ while that of $h_{n+1,n}$ is almost independent of it (as long as $\tilde{\mu} \lesssim 1$). For the values of $n$ accessible to us, the error $\varepsilon_n$ is much smaller for order-unity dissipation $\tilde{\mu} = 1$. We also observe a rather fast growth of $\varepsilon_n$ as $n$ increases, and cannot rule out the possibility that the matrix $h_{m,n}$ deviates significantly from the ideal scenario for $n$ sufficiently large. However, for weaker dissipation, the errors appear to be well described by $\varepsilon_n \propto \mu n$, with a proportionality constant of order $10^{-2}$. Thus, the matrix should be close to the ideal scenario for $n \lesssim 10^2 / \mu$. As we shall see below, the operator dynamics will be confined in the region $n \lesssim 1/\mu$ under the ideal scenario, so the deviation from the ideal scenario at larger $n$ is irrelevant.

To conclude, the large $N$ numerics confirms that operator size concentration is a good approximation at finite $q$, and as a consequence, the Arnoldi iteration results in a matrix close to the ``ideal scenario'': symmetric and tridiagonal, with diagonals $h_{n,n} \sim i \chi \mu n$ and primary off-diagonals $h_{n+1,n} \sim \alpha n$. 

\subsection{Arnoldi iteration at finite $N$}

\begin{figure}[t]
\subfigure[]{\includegraphics[height=4.7cm,width=0.47\linewidth]{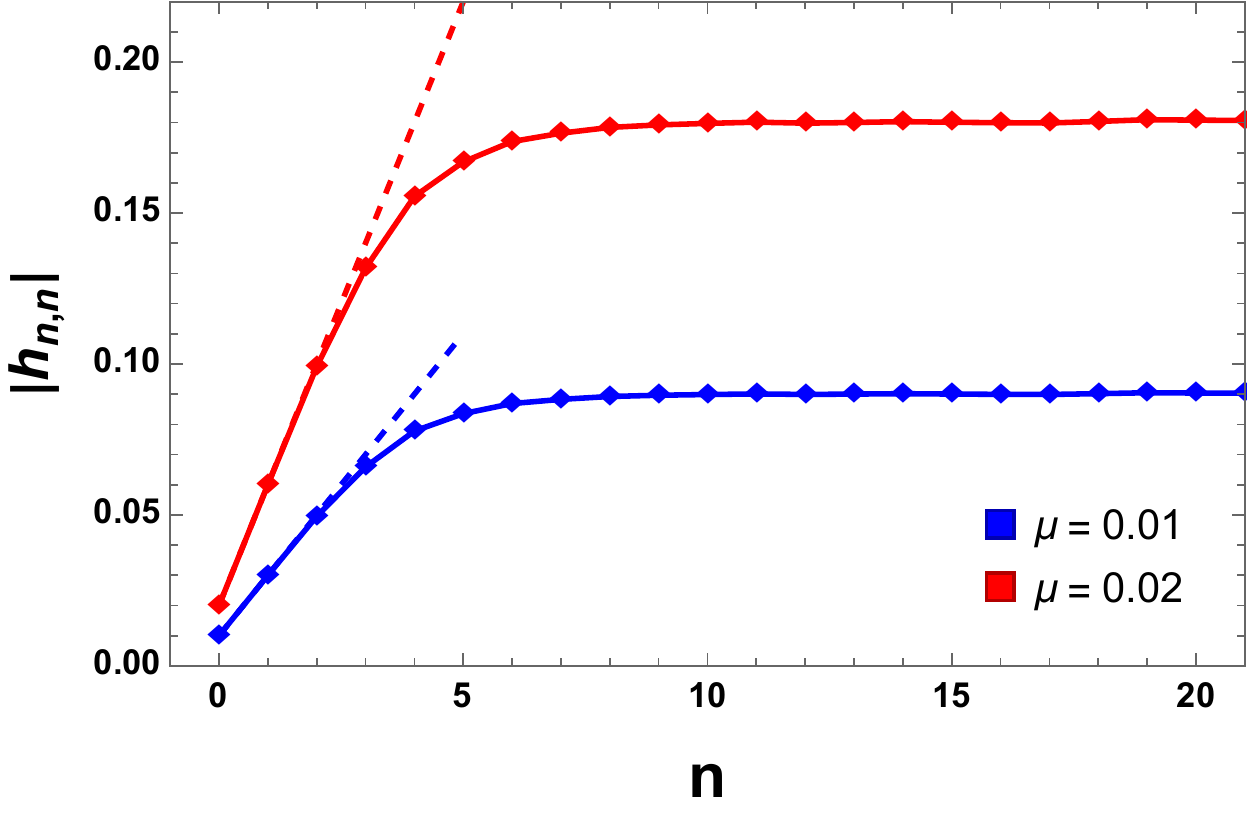}}
\hfill
\subfigure[]{\includegraphics[height=4.7cm,width=0.47\linewidth]{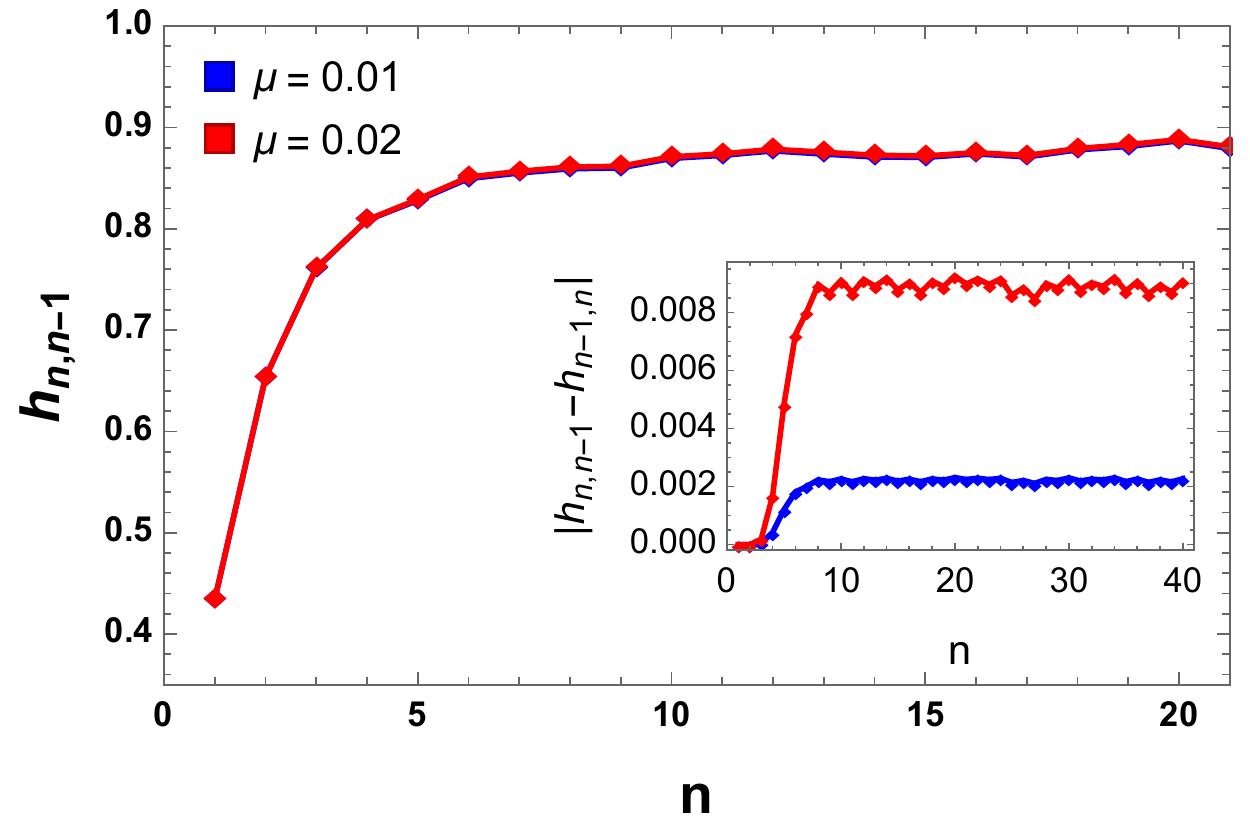}}
\hfill
\subfigure[]{\includegraphics[height=4.7cm,width=0.47\linewidth]{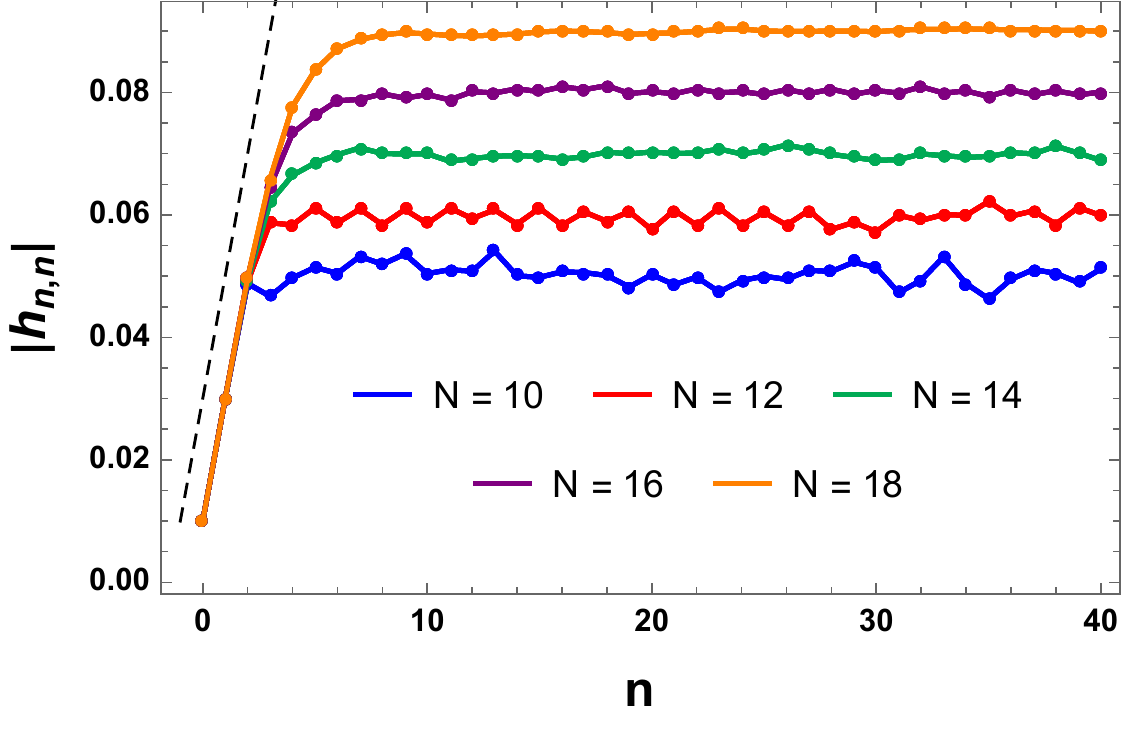}}
\hfill
\subfigure[]{\includegraphics[height=4.7cm,width=0.47\linewidth]{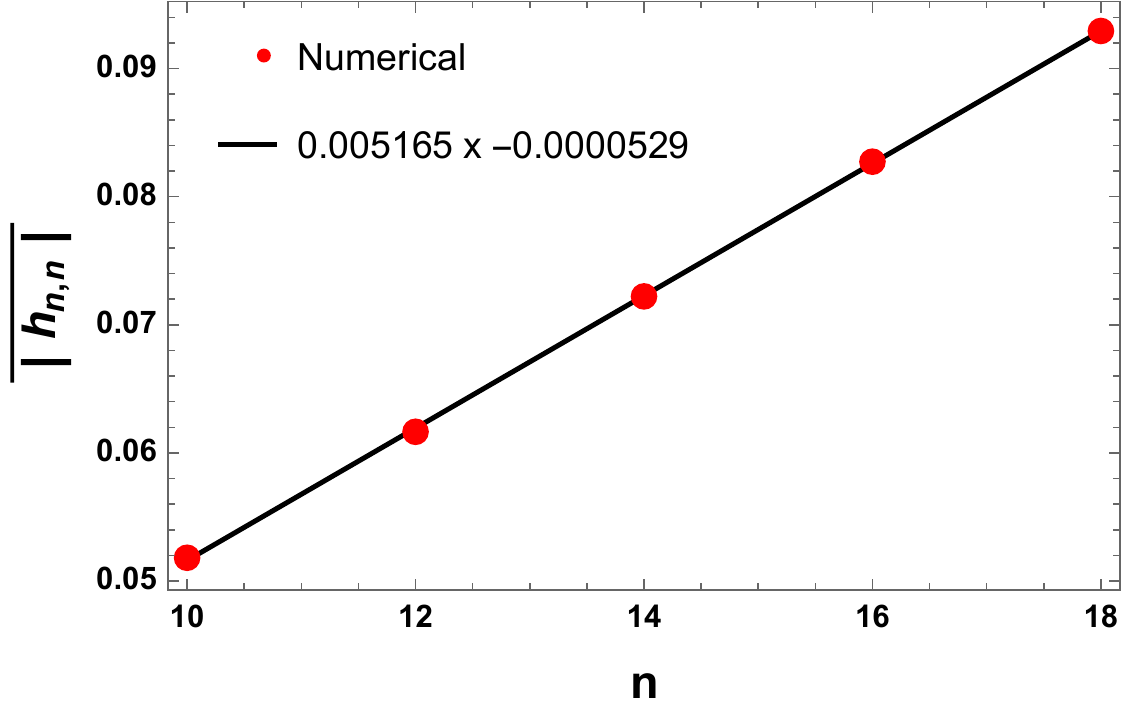}}
\caption{Behavior of the (a) primary diagonal and (b) off-diagonal elements respectively, for two different dissipation $\mu = 0.01$ and $\mu = 0.02$. The dotted lines in (a) are the fitted straight lines \eqref{fithnn}. The inset in (b) shows the difference $|h_{n,n-1} - h_{n-1,n}|$ grows with the dissipation. (c) The behavior of the diagonal elements (magnitude) for different system sizes. The black dotted line denotes the straight line \eqref{fithnn} (with an offset introduced for visualization). (d) The saturation value (averaged for $n=10$ to $n=40$) of the diagonal elements for different system sizes. In all cases, our system size is $N = 18$ (single realization) and we choose $q = 4$, $J=1$, and the fixed dissipation $\mu = 0.01$ in (c) and (d).} \label{fig:hnm}
\end{figure}

We now consider Arnoldi iteration for single realizations of the $q=4$ SYK model at finite $N$, up to $N = 18$, starting from the (normalized) initial operator $\sqrt{2} \psi_1$. A sample of the resulting matrix $[h_{m,n}]$ is plotted in Fig.~\ref{fig:arnoldi-syk4-anbn} (b). It is still close to the ideal scenario, modulo the saturation effect induced by the finite system size (see below). The diagonal elements $h_{n,n}$ are imaginary and grow linearly with $n$ up to the finite size saturation. The growth rate compares well with the prediction (see also Fig.~\ref{fig:hnm}(a)):
\begin{equation}
    h_{n,n} \approx i \mu s =  i \mu (2 n+1) \,. \label{fithnn}
\end{equation}  
In other words, $\chi = q-2 = 2$ in Eq.~\eqref{eq:hnn-ideal}. To obtain \eqref{fithnn},  we assume that operator size distribution is still a good approximation. Hence, we have $\mathcal{V}_n \approx \mathcal{O}_n$, and the operator size distribution of $\mathcal{V}_n$ is concentrated at $(q-2) n + 1 = 2n+1$. Then \eqref{fithnn} follows from recalling how the dissipative Lindbladian acts on Majorana strings~\eqref{eq:LDMajorana}. The primary off-diagonals are also not exactly equal, $h_{n,n-1} \neq h_{n-1,n}$, see Fig.~\ref{fig:hnm} (b), yet the relative difference is small. The magnitudes of the other off-diagonal elements are also small compared to diagonal and primary off-diagonal elements which are manifested in Fig.~\ref{fig:arnoldi-syk4-anbn} (b). 

Finally, in Fig.~\ref{fig:hnm} (c-d), we exhibit the system size ($N$) dependence of the diagonal elements. In Fig.~\ref{fig:hnm} (c), we see a saturation plateau after an initial linear growth for $ n \lesssim N / q$. The saturation value  increases linearly with $N$, see Fig.~\ref{fig:hnm} (d). The finite size saturation happens since the Majorana string lengths are upper bounded by $N$, by \eqref{eq:LDMajorana}. A similar behavior applies to the primary off-diagonal entries, as in the closed case.

\section{Operator growth and Krylov complexity} \label{krylov}

In the preceding sections, we showed that --- exactly at the large $N$ and large $q$ limit, to a good approximation beyond --- the dissipative SYK model displays operator size concentration.  As a result, the three approaches of Section~\ref{sec:openkrylovreview} coincide, and the Lindbladian acts  on the Krylov basis, as a symmetric tridiagonal matrix whose diagonal and primary off-diagonal elements have the following asymptotic growth: 
\begin{equation}\label{eq:anbn-krylov}
    a_n \simeq i \chi \mu  n  \,,\,~~ b_n \simeq \alpha n \,,
\end{equation}
This is also referred to as the ``ideal scenario''. The coefficient $\alpha$ is the same as in the closed system, whereas $\chi$ depends on how the size of the Krylov basis operator $\mathcal{O}_n$ grows with $n$. It may be a bit hasty to discuss how general it is. Even if we restrict to situations where the ideal scenario entails operator size concentration, it is in general unknown how the latter is respected in models away from large $N$; however, there is numerical evidence indicating that the ideal scenario remains a good approximation in quantum spin chains~\cite{Bhattacharya:2022gbz, Liu:2022god}. We leave a detailed study to the future.

Nevertheless, we can readily examine the consequence of the ideal scenario on the K-complexity growth. For this, we expand the operator on the closed-system Krylov basis:
\begin{align}
    |\mathcal{O}(t)) = \sum_{n} i^n \varphi_n (t) |\mathcal{O}_n)\,.
\end{align}
Note that  the norm of the operator  is not conserved \cite{Liu:2022god}: $ \sum_{n} |\varphi_n (t)|^2 \ne 1$. Thus, we shall define the K-complexity as the average position $n$ of the normalized wavefunction:
\begin{equation}
   K(t) = \frac1{\mathcal{Z}} \sum_n n  |\varphi_n|^2 \,,\, ~~
   \mathcal{Z} := \sum_n  |\varphi_n|^2 \,.
\end{equation}
Under the ideal scenario, the amplitudes satisfy the following differential equation
\begin{align}
    \partial_t \varphi_n (t) = i a_n \varphi_n(t) - b_{n+1} \varphi_{n+1}(t) + b_{n} \varphi_{n-1}(t)\,. \label{phieq}
\end{align}
where $a_n$ and $b_n$ satisfy \eqref{eq:anbn-krylov}, and initially, the wavefunction is localized at the origin: $\varphi_n (t=0) = \delta_{n,0}. $

\subsection{General argument}
To understand the asymptotic quantitative behavior in a simple way, we can make the Ansatz that $\varphi_n = \varphi(n)$ depends on $n$ smoothly, and approximate \eqref{phieq} by a PDE:
\begin{equation}
     \partial_t \varphi + n ( \chi \mu  \varphi + 2 \alpha  \partial_n \varphi ) = 0 \,.
\end{equation}
This equation has a stationary ($t \to\infty$) solution:
\begin{equation}
     \varphi_*(n) \propto  e^{- n / \xi} \,,\, ~~\xi := \frac{2\alpha}{\chi \mu}\,. \label{eq:stationary}
\end{equation}
It has an exponential tail of width $\xi$, which is inverse proportional to the dissipation strength $\mu$, and proportional to the K-complexity growth rate in the closed system $2\alpha$. The continuum approach here is quantitatively correct if $\xi \gg 1$, i.e., in the weak dissipation regime. In this regime, the dissipation term is negligible for $n \ll \xi$.  Now, initially, the wavefunction is localized at origin, so it will evolve as in the $\mu = 0$ case: it is known~\cite{Parker:2018yvk} that it will spread exponentially fast, and its average position $K(t) \sim e^{2\alpha t}$. This is true until the wavefunction reaches $K(t) \sim \xi$, where the dissipation term can no longer be neglected. In fact, it suppresses any further growth of the K-complexity, and the wavefunction converges to the stationary solution.\footnote{To be precise, this is true modulo the global normalization, which decays exponentially in the stationary regime; the decay rate depends on details at $n$ small. But at $n$ large, the spatial profile is still given by \eqref{eq:stationary}.} 

We have thus argued that, under the ideal scenario and in the weak dissipation regime, the exponential growth of the K-complexity saturates at $\xi$:
\begin{equation} \label{eq:Ktgen}
    K(t) \sim \begin{cases}
        e^{2\alpha t} & t \lesssim t_* \\
        \xi   & t \gtrsim t_*
    \end{cases}  \,,\,~~ t_* \sim \frac1{2\alpha} \ln \left( \frac{2\alpha}{\chi \mu} \right) \,.
\end{equation}
Here we also identified the saturation time scale, which diverges as a log of the inverse dissipation strength as $\mu \to 0$ --- a behavior reminiscent of the scrambling time with respect to a finite system. At strong dissipation, $t_* \to 1$ and $\xi \to 1$, the exponential growth regime of $K(t)$ disappears. 

It is interesting to note that, the ``localization'' of the operator wavefunction in the region $n \lesssim \xi \propto \alpha/\mu$ on the Krylov chain protects itself against any deviation from the ideal scenario, provided that happens only at $n \gg \xi$; We have seen in Section~\ref{sec:largeNarnoldi} that this is indeed the case for the dissipative SYK at finite $q$. It is reasonable to expect this to be general, as the deviation from the ideal scenario necessitates dissipation. Therefore, we believe the K-complexity saturation to be a robust phenomenon in similar situations.

\begin{figure}[t]
\subfigure[]{\includegraphics[height=4.6cm,width=0.47\linewidth]{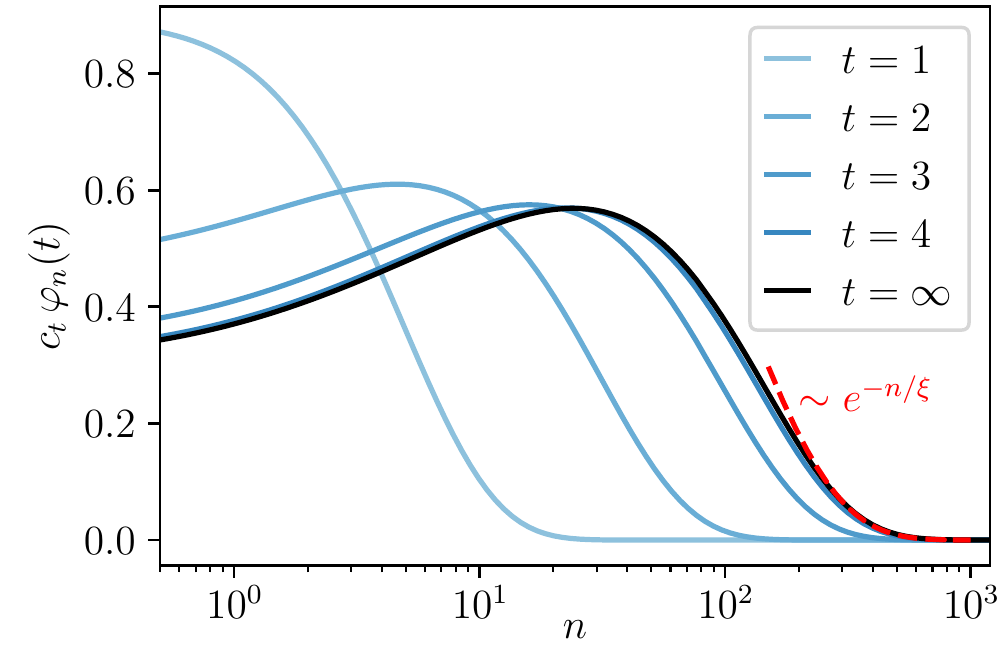}}
\hfill
\subfigure[]{\includegraphics[height=4.8cm,width=0.47\linewidth]{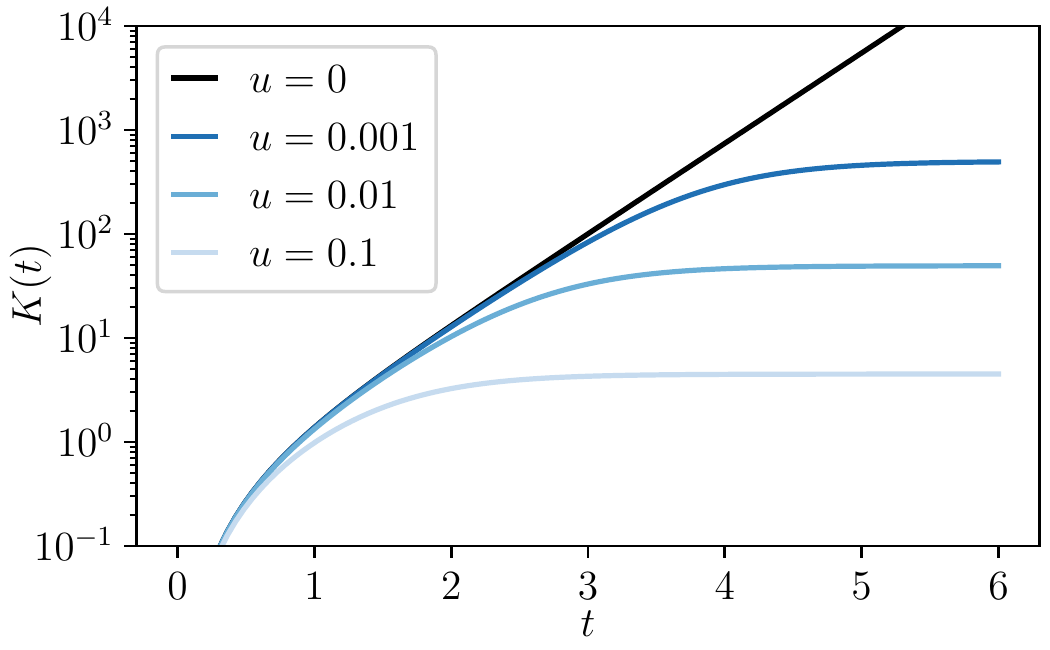}}
\caption{(a) Snapshots of the wave-function on the Krylov chain, from the exact solution \eqref{eq:anbnexact}, \eqref{eq:phin_exact} with $\eta = 1.5$ and $u = 0.01$. For display, we rescaled the wavefunctions by a $t$-dependent constant; we also interpolated $\varphi_n$ to non-integer values of $n$. At $t\to\infty$, a stationary profile is reached with an exponential tail $\propto e^{-n/\xi}$ with $\xi = 1/u$, indicated by the dashed curve. (b) Growth of the K-complexity for various dissipation strength ($u$), from the exact solution \eqref{eq:Ktex}. The exponential growth, which lasts indefinitely in the closed system ($u=0$), saturates at $t \sim \ln (1/u)$, or $K(t) \sim 1/u$, in the presence of dissipation.} \label{fig:wave}
\end{figure}

\subsection{Exact solution}
The above argument, although heuristic, is applicable to any $a_n$ and $b_n$ with the asymptotics \eqref{eq:anbn-krylov}. Here, we check the conclusion with an exact solvable set of coefficients:
\begin{equation}\label{eq:anbnexact}
    b_n^2 = (1 - u^2) n (n-1+\eta) \,,\, ~~a_n = i u (2 n  + \eta) \,.
\end{equation}
This is precisely the recursion coefficients of the Meixner polynomials used in Ref.~\cite{Parker:2018yvk} (Appendix D, Eq.~(D10), upon identifying $\delta = i u$; in that work, only $\delta = 0$ was useful). These Lanczos coefficients have the asymptotic behavior \eqref{eq:anbn-krylov}, with
\begin{equation}\label{eq:meixner-param} \alpha = 1 - u^2 \,,\,~~ \chi \mu = 2 u \,. \end{equation}
Varying $u \in (0,1)$, we have obtained any relative dissipation strength, $\xi^{-1} =  \chi \mu /2 \alpha  \in (0,\infty)$ (meanwhile $\eta > 0$ does not affect the asymptotic growth rates, but only the offsets).

Following closely \textit{ibid.} (Eq D11 and D9), it is straightforward to show that 
\begin{align} \label{eq:phin_exact}
   \varphi_n(t) =   \frac{ \sech(t)^\eta }{(1 + u \tanh(t))^\eta} \,\times 
     (1 - u^2)^{\frac{n}2}  \sqrt{\frac{(\eta)_n}{n!}}    \left( \frac{\tanh (t)}{1 +u \tanh(t)} \right)^n \,.
\end{align}
Here $(\eta)_n = \eta (\eta + 1) \dots (\eta + n-1) $ is the Pochhammer symbol. Since we will use the normalized wavefunction to define the K-complexity, the first line of \eqref{eq:phin_exact} (an $n$ independent pre-factor) can be ignored. In Fig.~\ref{fig:wave} (a), we plotted a few wavefunction snapshots from the above exact solution for $\eta = 1.5$. We see qualitatively that time evolution is indeed described by the above general argument. Curiously, the coefficients \eqref{eq:anbnexact} have the same structure of $\mathrm{SL}(2, \mathbb{R})$, after identifying the coefficients \cite{Balasubramanian:2022tpr} and the highest weight state. Hence, the amplitude \eqref{eq:phin_exact} is fixed by the $\mathrm{SL}(2, \mathbb{R})$ symmetry, which possibly controls the complexity growth \cite{Hornedal:2022pkc}. As we see, in our case, the $\mathrm{SL}(2, \mathbb{R})$ structure is preserved, yet the complexity is suppressed. The suppression entirely comes due to the imaginary part of the coefficients $a_n$.\footnote {We thank Pawel Caputa for pointing this out.} 

We can make further quantitative comparisons, focusing on the tail of the wave function. It is not hard to see that at $t\to\infty$, the wavefunction has the following stationary $n$-dependence for $u > 0$:
\begin{align}
    \varphi_n(t\to\infty) \simeq \left( \frac{\sqrt{1-u^2}}{1+u} \right)^n  n^{\frac{\eta-1}2} \,.
\end{align}
We find an exponential decay.\footnote{The power-law correction comes from $\sqrt{(\eta)_n / n!}$, after using the asymptotics $\Gamma(z+p)/\Gamma(z+q) \sim z^{p-q}$ as $z \rightarrow \infty$.} At small $u$ (weak dissipation), its inverse width 
$$  \ln \left( \frac{1+u}{\sqrt{1-u^2}}  \right) = u + O(u^3)\,, $$
agrees with the prediction of \eqref{eq:stationary} applied to \eqref{eq:meixner-param}, which gives also $\xi^{-1} = \chi \mu / (2\alpha) = u +  O(u^3)$. A similar analysis at finite $t$ gives
\begin{equation}
    \varphi_n(t) \sim e^{-n / \xi(u,t)} n^{\frac{\eta-1}2}\,,
\end{equation}
where 
\begin{equation}
    \xi(u,t)^{-1} = u+2e^{-2 t} + O(e^{-4t}, e^{-2t} u, u^2) \,.
\end{equation}
The saturation time is given by equating the two terms; we thus find $t = \ln (1/u)/ 2 + O(1)$ which agrees with \eqref{eq:Ktgen} at small $u$.

The exact solution above allows us to compute the K-complexity analytically. Indeed, denoting 
\begin{equation}
    e^{y_0} = (1 - u^2) \left( \frac{\tanh (t)}{1 + u \tanh(t)} \right)^{2} \,,
\end{equation}
it is not hard to see from \eqref{eq:phin_exact} that 
\begin{align}
    K(t) = \frac{1}{\mathcal{Z}} \sum_n  n \, |\varphi_n (t)|^2  
    &= \left.\partial_y \ln \left( \sum_n e^{y n} \frac{(\eta)_n}{n!} \right)\right|_{y = y_0}  \nonumber \\
    &=  \left.\partial_y \ln \left((1- e^{y})^{-\eta} \right)  \right|_{y = y_0} \nonumber \\
    &= \frac{\eta\left(1-u^2\right) \tanh ^2(t)}{1+ 2 u \tanh (t)- \left(1-2 u^2\right) \tanh ^2(t)} \,. \label{eq:Ktex}
\end{align}
See Fig.~\ref{fig:wave} (b) for a plot. Expanding this around $u = 0$ (weak dissipation) at fixed $t$ gives
\begin{equation}
    K(t) = \eta \left[ \sinh ^2(t)-2 u  \sinh ^3(t) \cosh (t) + O(u^2)\right] \,,
\end{equation}
where we recover the exponential growth $K(t) \sim e^{2\alpha t}$ (with $\alpha = 1$) in the closed system, as well as the  first correction due to dissipation. Now, the latter will dominate at $t \sim t_* \sim \ln (1/u) /2 $, rendering this expansion useless. Meanwhile, expanding around $t\to\infty$ at fixed $u>0$,  we see that the K-complexity goes to a constant:
\begin{equation}
    K(t\to\infty) = \frac{\eta}{2 u} - \frac{\eta}{2}  \,.
\end{equation}
It is proportional to $\xi \sim 1/u$ at small $u$, and the proportionality constant is controlled by $\eta$. Hence, the exact calculation confirms the conclusions~\eqref{eq:Ktgen} of the general argument.

Similarly, we can compute the normalized variance~\cite{Caputa:2021ori, Bhattacharjee:2022ave} (note that we define it again using the normalized wavefunction):
\begin{align}
\left( n^2 \right)_c :=\,  \frac{1}{\mathcal{Z}} \sum_n |\varphi_n (t)|^2 (n - K(t))^2  
=&\left.\partial_y^2 \ln \left( \sum_n e^{y n} \frac{(\eta)_n}{n!} \right)\right|_{y = y_0} \nonumber \\
=& \,\frac{\eta\left(1-u^2\right) \tanh ^2(t) (u \tanh (t)+1)^2}{\left(1+ 2 u \tanh (t)- \left(1-2 u^2\right) \tanh ^2(t)\right)^2} \,.
\end{align}
It behaves as $\eta e^{4t}$ in the growth regime and saturates at $\eta / (4u^2)$ in the late time regime. In both cases, the position fluctuation $\sqrt{\left( n^2 \right)_c}$ is comparable to its average $K(t)$.

It is interesting to remark that the Lanczos coefficient \eqref{eq:an-main},\eqref{eq:bn-main} of the large $q$ SYK corresponds exactly --- up to subleading terms in a $1/q$ expansion --- to the exact solvable ones \eqref{eq:bn-main} upon identifying 
\begin{equation} \eta = 2/q \,,\,~~ 2 u =  \tilde\mu  \,,\,~~ 
\mathcal{J}^2 = 1 - u^2 \,. \label{eq:sykqexact} \end{equation}
As such, our results above apply to the large $q$ SYK; in particular, we may recover \eqref{eq:Ct} and \eqref{eq:cfformula}. The K-complexity in this model is $\propto 1/q$, as most of the wave function is localized at the origin.

\subsection{Bound on chaos}
We now come back to discussing a general consequence of the saturated K-complexity growth \eqref{eq:Ktgen}. Now, under the ideal scenario, the K-complexity is computed with respect to the closed-system Krylov basis (which is also approximately the basis generated by the Arnoldi iteration). Thus, the argument leading to the bound on chaos in Ref.~\cite{Parker:2018yvk} still applies and entails that the exponential growth of the normalized out-of-time-order correlator (OTOC) must saturate. More precisely and concretely, in the dissipative SYK model, we expect (and have proven at large $q$) that there exists a $t$-independent constant such that 
\begin{equation}\label{eq:OTOC}
  \frac{1}{\mathcal{Z}} \sum_{j=1}^N \mathrm{Tr}[\rho_{\infty}\left\{ \psi_j ,\psi_1(t) \right\} \left\{ \psi_j ,\psi_1(t) \right\}^\dagger] \le C K(t)\,,
\end{equation}
where $K(t)$ is given by \eqref{eq:Ktgen}, with $\chi$ and $\alpha$ calculable from the model parameters, and $\mathcal{Z} = \mathrm{Tr}[\rho_{\infty} \psi_1(t)^\dagger \psi_1(t)] $ is the norm of the evolved operator. In particular, in the presence of any dissipation, however small it is, the OTOC at the left-hand side cannot grow indefinitely even in the $N\to\infty$ limit. In particular, there is no nonzero critical value of $\mu_c$ below which the OTOC is in a ``chaotic'' phase, in contrast to the models studied recently~\cite{Weinstein:2022yce, Zhang:2022knu}. Indeed, the dissipation term in our case has a more severe effect on the terms in $ \psi_1(t)$ with large size (long Majorana strings), which dominate the OTOC. They are simply eliminated (with a rate proportional to the length); meanwhile, in the models of \textit{op. cit.}, the dissipation reduces their sizes without eliminating them altogether.  


\subsection{Holographic interpretation}
Finally, we speculate some heuristic explanations of our results from the dual gravity picture. As argued in \cite{Garcia-Garcia:2022adg}, the SYK model with the jump operators given by \eqref{jumpop} is dual to a Keldysh wormhole. The gravitational geometry of such a wormhole is far from clear. The configuration is valid in real-time and is fundamentally different from the traversable \cite{Maldacena:2018lmt} or Euclidean wormhole \cite{Garcia-Garcia:2020ttf}.  However, we can still take inspiration from the bulk picture of the growth of operator size \cite{Susskind:2018tei, Susskind:2020gnl}. The scrambling time in the dissipative SYK appears to be smaller than in the non-dissipative SYK. Increasing the coupling decreases the scrambling time, encountering a possible competition between the scrambling and decoherence \cite{Xu:2020wky}. In the dual picture, we might think that the particle falling on one side can be reached on the other side due to the coupling like Eq.~\eqref{haam}.  However, there is still a finite decay rate even at zero dissipation \cite{Kulkarni:2021xsx, Garcia-Garcia:2022adg}, which is possibly related to the length of the wormhole. The dissipation effectively increases the coupling between the two sides of the wormhole, thus signaling a quicker scrambling. It will be interesting to understand this in a much broader sense, especially computing the complexity in the Jackiw-Teitelboim (JT) gravity \cite{Jian:2020qpp}. We leave a detailed discussion in future studies.

\section{Conclusion} \label{concl}

In this paper, we aim to extend the universal operator growth hypothesis in open quantum systems. Our model of study is the dissipative $q$-body SYK model where the interactions with the environment are modeled by the Lindbladian construction. One of the particular motivations is the large $q$ limit, where the quantities of interest can be computed analytically. In particular, the ``operator size concentration'' allows a simple proof to extract two sets of Lanczos coefficients, which can also be exactly computed from the moment method. Out of two sets, only one of the sets is sensitive to dissipation while the other keeps track of the chaotic nature of the Hamiltonian. The asymptotic growth of both sets of coefficients is linear, which can be considered a generalized operator growth hypothesis in the dissipative systems. This result is consistent with the semi-analytics in $q$ by solving the Schwinger-Dyson equations and explicit numerical implementation of Arnoldi iteration. The dissipative nature of the environment suppresses the complexity compared to its exponential growth in a closed system. This result corroborates the finds of SYK for finite $N$ (thus finite $q$) and spinless fermionic systems \cite{Liu:2022god}. We believe that the exponential growth and the consecutive saturation at ``scrambling time'' is a universal feature to hold any dissipative systems. Finally, we argue some holographic interpretation of the complexity growth of the dissipative SYK, known to be dual to the Keldysh wormhole \cite{Garcia-Garcia:2022adg}. 

While the asymptotic growth of $ b_n$s has been known for integrable systems, it is an open question to ask about the asymptotic growth of the $a_n$s in generic integrable systems. Some relevant studies have been performed in \cite{Bhattacharya:2022gbz}, and the growth seems to be not affected due to the integrability of the system. We believe a careful study is required. It will also be worth implementing the numerical bi-Lanczos algorithm \cite{gruning}, where the ``bra'' and ``ket'' vectors in the doubled Hilbert space will evolve differently. The Lindbladian in this bi-orthonormal basis is expected to reproduce the analytical results in the large $N$ and large $q$ limit. Finally, it is also exciting to see the growth of the complexity with random jump operators \cite{Sa:2021tdr}, and especially for the dissipative dynamics of complex fermions, where the latter might be equivalent to two non-Hermitian SYK \cite{Garcia-Garcia:2022rtg, Cai:2022onu}. This might provide a general perspective to establish a ``universal'' conclusion of our results. A larger and wider motivation is to understand the chaos bound \cite{Maldacena:2015waa} in dissipative systems, provided it is well-defined in such a scenario. We hope to return to some of these questions in future works.

\section{Acknowledgements}

X.C. thanks Thomas Scaffidi and Zihao Qi for useful discussions, in particular for pointing out Ref.~\cite{Liu:2022god}. P.N. thanks Pawel Caputa, Tokiro Numasawa, and Shinsei Ryu for the fruitful discussions. P.N. would like to thank the organizers of Kavli Asian Winter School (KAWS) 2023 at Daejeon, Korea, and the young researchers' workshop of Extreme Universe (ExU) collaboration at Nagoya University, Japan, where the results were presented. B.B. is supported by the Ministry of Human Resource Development, Government of India through the Prime Ministers' Research Fellowship. The work of P.N. is supported by the JSPS Grant-in-Aid for Transformative Research Areas (A) ``Extreme Universe'' No. 21H05190.

\appendix

\section{Appendix: Derivation of Eq.~\eqref{eq:LDMajorana}} \label{appa}
In this Appendix, we outline a derivation of Eq.~\eqref{eq:LDMajorana}. Notice that our jump operators \eqref{jumpop} are always fermionic. Depending on the string operators, we have the following two cases:\\
\newline
\textbf{Case 1:} When the string operators are fermionic, we have to use the negative sign in the Lindbladian \eqref{negpos}. In this case, the string of operators has to be even to be a fermionic operator. With the jump operators \eqref{jumpop}, we have
\begin{align}
    \mathcal{L}_D \psi_{i_1} \cdots \psi_{i_s} = i \mu\sum_{k = 1}^N \psi_k (\psi_{i_1} \cdots \psi_{i_s}) \psi_k
    + \frac{i \mu}{2} \sum_{k = 1}^N \bigg\{\frac{I}{2}, \psi_{i_1} \cdots \psi_{i_s}\bigg\}\,,
\end{align}
for odd $s$. Here we have used $\psi_k^{\dagger} = \psi_k$ and $\psi_k^2 = I/2$, where $I$ is the identity matrix. The second summation is easier to compute and it gives
\begin{align}
    \frac{i \mu}{2} \sum_{k = 1}^N \bigg\{\frac{I}{2}, \psi_{i_1} \cdots \psi_{i_s}\bigg\} = \frac{i \mu N}{2} \psi_{i_1} \cdots \psi_{i_s}\,. \label{easyodd}
\end{align}
The first summation is more involved. To simply it, we break it into two parts
\begin{align*}
    i \mu\sum_{k = 1}^{N} \psi_k (\psi_{i_1} \cdots \psi_{i_s}) \psi_k = i \mu\sum_{k = 1}^{i_s} \psi_k (\psi_{i_1} \cdots \psi_{i_s}) \psi_k 
    + i \mu\sum_{k = i_{s+1}}^N \psi_k (\psi_{i_1} \cdots \psi_{i_s}) \psi_k\,. 
\end{align*}
Now the first sum is up to the string length $s$. It requires $(s-1)$ shifts to rearrange according to the string length. On the other hand, the second sum involves $(N-s)$ terms and requires $s$ shifts. Hence, the result is
\begin{align*}
    i \mu\sum_{k = 1}^{N} \psi_k (\psi_{i_1} \cdots \psi_{i_s}) \psi_k &= \frac{(-1)^{s-1}}{2} i \mu s\, \psi_{i_1} \cdots \psi_{i_s} + \frac{(-1)^{s}}{2} i \mu (N-s)\, \psi_{i_1} \cdots \psi_{i_s}\,, \nonumber \\
    & = \frac{(-1)^{s}}{2} i \mu (N-2s)\, \psi_{i_1} \cdots \psi_{i_s}\,.
\end{align*}
Since $s$ is odd, we readily have
\begin{align*}
    i \mu\sum_{k = 1}^{N} \psi_k (\psi_{i_1} \cdots \psi_{i_s}) \psi_k = -\frac{1}{2} i \mu (N-2s)\, \psi_{i_1} \cdots \psi_{i_s}\,.
\end{align*}
Combining with \eqref{easyodd}, we immediately get \eqref{eq:LDMajorana}. Notice that, as $N$ is even, the RHS of the above equation cannot be zero, it can be positive or negative depending on the number of fermions and the string length.\\
\newline
\textbf{Case 2:} When the string operator is bosonic, we have to use the plus sign in the Lindbladian \eqref{negpos}. In this case, the string of operators has to be even to be a bosonic operator. With the same jump operators \eqref{jumpop}, now we have
\begin{align}
    \mathcal{L}_D \psi_{i_1} \cdots \psi_{i_s} &= -i \mu\sum_{k = 1}^N \psi_k (\psi_{i_1} \cdots \psi_{i_s}) \psi_k 
    + \frac{i \mu}{2} \sum_{k = 1}^N \bigg\{\frac{I}{2}, \psi_{i_1} \cdots \psi_{i_s}\bigg\}\,, \label{bos}
\end{align}
for even $s$. The previous computations will exactly follow. However, now we have
\begin{align}
    i \mu\sum_{k = 1}^{N} \psi_k (\psi_{i_1} \cdots \psi_{i_s}) \psi_k = \frac{1}{2} i \mu (N-2s)\, \psi_{i_1} \cdots \psi_{i_s}\,. \label{zerr}
\end{align}
Moreover, the Eq.~\eqref{easyodd} will still hold for even $s$. Hence, combining \eqref{easyodd} with \eqref{bos}, we find \eqref{eq:LDMajorana}. Notice that, when $N=2s$, the first summation in \eqref{bos} will vanish as evident from Eq.~\eqref{zerr}. This can be straightforwardly checked by choosing a particular $N$ and $s$.

\bibliographystyle{JHEP}
\bibliography{references}  

\providecommand{\href}[2]{#2}\begingroup\raggedright\begin{thebibliography}{10}

\bibitem{Parker:2018yvk}
D.~E. Parker, X.~Cao, A.~Avdoshkin, T.~Scaffidi, and E.~Altman, {\it {A
  Universal Operator Growth Hypothesis}},  {\em Phys. Rev. X} {\bf 9} (2019),
  no.~4 041017, [\href{http://arxiv.org/abs/1812.08657}{{\tt
  arXiv:1812.08657}}].

\bibitem{Breuer2007}
H.-P. Breuer and F.~Petruccione, {\em The Theory of Open Quantum Systems}.
\newblock Oxford University Press, Oxford, 2007.

\bibitem{Lindblad1976}
G.~Lindblad, {\it On the generators of quantum dynamical semigroups},  {\em
  Communications in Mathematical Physics} {\bf 48} (Jun, 1976) 119--130.

\bibitem{Gorini}
V.~Gorini, A.~Kossakowski, and E.~C.~G. Sudarshan, {\it Completely positive
  dynamical semigroups of n‐level systems},  {\em Journal of Mathematical
  Physics} {\bf 17} (1976), no.~5 821--825,
  [\href{http://arxiv.org/abs/https://aip.scitation.org/doi/pdf/10.1063/1.522979}{{\tt
  https://aip.scitation.org/doi/pdf/10.1063/1.522979}}].

\bibitem{PhysRevLett.61.1899}
R.~Grobe, F.~Haake, and H.-J. Sommers, {\it Quantum distinction of regular and
  chaotic dissipative motion},  {\em Phys. Rev. Lett.} {\bf 61} (Oct, 1988)
  1899--1902.

\bibitem{PhysRevLett.123.254101}
G.~Akemann, M.~Kieburg, A.~Mielke, and T.~Prosen, {\it Universal signature from
  integrability to chaos in dissipative open quantum systems},  {\em Phys. Rev.
  Lett.} {\bf 123} (Dec, 2019) 254101,
  [\href{http://arxiv.org/abs/1910.03520}{{\tt arXiv:1910.03520}}].

\bibitem{Xu:2020wky}
Z.~Xu, A.~Chenu, T.~Prosen, and A.~del Campo, {\it {Thermofield dynamics:
  Quantum Chaos versus Decoherence}},  {\em Phys. Rev. B} {\bf 103} (2021),
  no.~6 064309, [\href{http://arxiv.org/abs/2008.06444}{{\tt
  arXiv:2008.06444}}].

\bibitem{Li:2021kuv}
J.~Li, T.~Prosen, and A.~Chan, {\it {Spectral Statistics of Non-Hermitian
  Matrices and Dissipative Quantum Chaos}},  {\em Phys. Rev. Lett.} {\bf 127}
  (2021), no.~17 170602, [\href{http://arxiv.org/abs/2103.05001}{{\tt
  arXiv:2103.05001}}].

\bibitem{Garcia-Garcia:2021rle}
A.~M. Garc\'\i{}a-Garc\'\i{}a, L.~S\'a, and J.~J.~M. Verbaarschot, {\it
  {Symmetry Classification and Universality in Non-Hermitian Many-Body Quantum
  Chaos by the Sachdev-Ye-Kitaev Model}},  {\em Phys. Rev. X} {\bf 12} (2022),
  no.~2 021040, [\href{http://arxiv.org/abs/2110.03444}{{\tt
  arXiv:2110.03444}}].

\bibitem{Matsoukas-Roubeas:2022odk}
A.~S. Matsoukas-Roubeas, F.~Roccati, J.~Cornelius, Z.~Xu, A.~Chenu, and A.~del
  Campo, {\it {Non-Hermitian Hamiltonian deformations in quantum mechanics}},
  {\em JHEP} {\bf 01} (2023) 060, [\href{http://arxiv.org/abs/2211.05437}{{\tt
  arXiv:2211.05437}}].

\bibitem{Kawabata:2022cpr}
K.~Kawabata, A.~Kulkarni, J.~Li, T.~Numasawa, and S.~Ryu, {\it {Symmetry of
  open quantum systems: Classification of dissipative quantum chaos}},
  \href{http://arxiv.org/abs/2212.00605}{{\tt arXiv:2212.00605}}.

\bibitem{Zhang:2022knu}
P.~Zhang and Z.~Yu, {\it {Dynamical Transition of Operator Size Growth in Open
  Quantum Systems}},  \href{http://arxiv.org/abs/2211.03535}{{\tt
  arXiv:2211.03535}}.

\bibitem{Omanakuttan:2022ikz}
S.~Omanakuttan, K.~Chinni, P.~D. Blocher, and P.~M. Poggi, {\it {Scrambling and
  quantum chaos indicators from long-time properties of operator
  distributions}},  \href{http://arxiv.org/abs/2211.15872}{{\tt
  arXiv:2211.15872}}.

\bibitem{pfz}
P.~Zhang and Y.~Gu, {\it {Operator Size Distribution in Large $N$ Quantum
  Mechanics of Majorana Fermions}},
  \href{http://arxiv.org/abs/2212.04358}{{\tt arXiv:2212.04358}}.

\bibitem{PhysRevA.103.062214}
P.~Zanardi and N.~Anand, {\it Information scrambling and chaos in open quantum
  systems},  {\em Phys. Rev. A} {\bf 103} (Jun, 2021) 062214,
  [\href{http://arxiv.org/abs/2012.13172}{{\tt arXiv:2012.13172}}].

\bibitem{Schuster:2022bot}
T.~Schuster and N.~Y. Yao, {\it {Operator Growth in Open Quantum Systems}},
  \href{http://arxiv.org/abs/2208.12272}{{\tt arXiv:2208.12272}}.

\bibitem{Weinstein:2022yce}
Z.~Weinstein, S.~P. Kelly, J.~Marino, and E.~Altman, {\it {Scrambling
  Transition in a Radiative Random Unitary Circuit}},
  \href{http://arxiv.org/abs/2210.14242}{{\tt arXiv:2210.14242}}.

\bibitem{Syzranov:2017zyp}
S.~V. Syzranov, A.~V. Gorshkov, and V.~Galitski, {\it {Out-of-time-order
  correlators in finite open systems}},  {\em Phys. Rev. B} {\bf 97} (2018),
  no.~16 161114, [\href{http://arxiv.org/abs/1704.08442}{{\tt
  arXiv:1704.08442}}].

\bibitem{Bhattacharya:2022gbz}
A.~Bhattacharya, P.~Nandy, P.~P. Nath, and H.~Sahu, {\it {Operator growth and
  Krylov construction in dissipative open quantum systems}},  {\em JHEP} {\bf
  12} (2022) 081, [\href{http://arxiv.org/abs/2207.05347}{{\tt
  arXiv:2207.05347}}].

\bibitem{Liu:2022god}
C.~Liu, H.~Tang, and H.~Zhai, {\it {Krylov Complexity in Open Quantum
  Systems}},  \href{http://arxiv.org/abs/2207.13603}{{\tt arXiv:2207.13603}}.

\bibitem{Rabinovici:2020ryf}
E.~Rabinovici, A.~S\'anchez-Garrido, R.~Shir, and J.~Sonner, {\it {Operator
  complexity: a journey to the edge of Krylov space}},  {\em JHEP} {\bf 06}
  (2021) 062, [\href{http://arxiv.org/abs/2009.01862}{{\tt arXiv:2009.01862}}].

\bibitem{Jian:2020qpp}
S.-K. Jian, B.~Swingle, and Z.-Y. Xian, {\it {Complexity growth of operators in
  the SYK model and in JT gravity}},  {\em JHEP} {\bf 03} (2021) 014,
  [\href{http://arxiv.org/abs/2008.12274}{{\tt arXiv:2008.12274}}].

\bibitem{Barbon:2019wsy}
J.~L.~F. Barb\'on, E.~Rabinovici, R.~Shir, and R.~Sinha, {\it {On The Evolution
  Of Operator Complexity Beyond Scrambling}},  {\em JHEP} {\bf 10} (2019) 264,
  [\href{http://arxiv.org/abs/1907.05393}{{\tt arXiv:1907.05393}}].

\bibitem{Dymarsky:2019elm}
A.~Dymarsky and A.~Gorsky, {\it {Quantum chaos as delocalization in Krylov
  space}},  {\em Phys. Rev. B} {\bf 102} (2020), no.~8 085137,
  [\href{http://arxiv.org/abs/1912.12227}{{\tt arXiv:1912.12227}}].

\bibitem{Cao:2020zls}
X.~Cao, {\it {A statistical mechanism for operator growth}},  {\em J. Phys. A}
  {\bf 54} (2021), no.~14 144001, [\href{http://arxiv.org/abs/2012.06544}{{\tt
  arXiv:2012.06544}}].

\bibitem{Dymarsky:2021bjq}
A.~Dymarsky and M.~Smolkin, {\it {Krylov complexity in conformal field
  theory}},  {\em Phys. Rev. D} {\bf 104} (2021), no.~8 L081702,
  [\href{http://arxiv.org/abs/2104.09514}{{\tt arXiv:2104.09514}}].

\bibitem{Caputa:2021sib}
P.~Caputa, J.~M. Magan, and D.~Patramanis, {\it {Geometry of Krylov
  complexity}},  {\em Phys. Rev. Res.} {\bf 4} (2022), no.~1 013041,
  [\href{http://arxiv.org/abs/2109.03824}{{\tt arXiv:2109.03824}}].

\bibitem{Rabinovici:2021qqt}
E.~Rabinovici, A.~S\'anchez-Garrido, R.~Shir, and J.~Sonner, {\it {Krylov
  localization and suppression of complexity}},  {\em JHEP} {\bf 03} (2022)
  211, [\href{http://arxiv.org/abs/2112.12128}{{\tt arXiv:2112.12128}}].

\bibitem{Bhattacharjee:2022vlt}
B.~Bhattacharjee, X.~Cao, P.~Nandy, and T.~Pathak, {\it {Krylov complexity in
  saddle-dominated scrambling}},  {\em JHEP} {\bf 05} (2022) 174,
  [\href{http://arxiv.org/abs/2203.03534}{{\tt arXiv:2203.03534}}].

\bibitem{Balasubramanian:2022tpr}
V.~Balasubramanian, P.~Caputa, J.~M. Magan, and Q.~Wu, {\it {Quantum chaos and
  the complexity of spread of states}},  {\em Phys. Rev. D} {\bf 106} (2022),
  no.~4 046007, [\href{http://arxiv.org/abs/2202.06957}{{\tt
  arXiv:2202.06957}}].

\bibitem{Hornedal:2022pkc}
N.~H\"ornedal, N.~Carabba, A.~S. Matsoukas-Roubeas, and A.~del Campo, {\it
  {Ultimate Physical Limits to the Growth of Operator Complexity}},  {\em
  Commun. Phys.} {\bf 5} (2022) 207,
  [\href{http://arxiv.org/abs/2202.05006}{{\tt arXiv:2202.05006}}].

\bibitem{Caputa:2021ori}
P.~Caputa and S.~Datta, {\it {Operator growth in 2d CFT}},  {\em JHEP} {\bf 12}
  (2021) 188, [\href{http://arxiv.org/abs/2110.10519}{{\tt arXiv:2110.10519}}].
  [Erratum: JHEP 09, 113 (2022)].

\bibitem{Bhattacharjee:2022ave}
B.~Bhattacharjee, P.~Nandy, and T.~Pathak, {\it {Krylov complexity in large-$q$
  and double-scaled SYK model}},  \href{http://arxiv.org/abs/2210.02474}{{\tt
  arXiv:2210.02474}}.

\bibitem{Bhattacharjee:2022qjw}
B.~Bhattacharjee, S.~Sur, and P.~Nandy, {\it {Probing quantum scars and weak
  ergodicity breaking through quantum complexity}},  {\em Phys. Rev. B} {\bf
  106} (2022), no.~20 205150, [\href{http://arxiv.org/abs/2208.05503}{{\tt
  arXiv:2208.05503}}].

\bibitem{Balasubramanian:2022dnj}
V.~Balasubramanian, J.~M. Magan, and Q.~Wu, {\it {A Tale of Two Hungarians:
  Tridiagonalizing Random Matrices}},
  \href{http://arxiv.org/abs/2208.08452}{{\tt arXiv:2208.08452}}.

\bibitem{Rabinovici:2022beu}
E.~Rabinovici, A.~S\'anchez-Garrido, R.~Shir, and J.~Sonner, {\it {Krylov
  complexity from integrability to chaos}},  {\em JHEP} {\bf 07} (2022) 151,
  [\href{http://arxiv.org/abs/2207.07701}{{\tt arXiv:2207.07701}}].

\bibitem{He:2022ryk}
S.~He, P.~H.~C. Lau, Z.-Y. Xian, and L.~Zhao, {\it {Quantum chaos, scrambling
  and operator growth in $ T\overline{T} $ deformed SYK models}},  {\em JHEP}
  {\bf 12} (2022) 070, [\href{http://arxiv.org/abs/2209.14936}{{\tt
  arXiv:2209.14936}}].

\bibitem{Guo:2022hui}
S.~Guo, {\it {Operator growth in SU(2) Yang-Mills theory}},
  \href{http://arxiv.org/abs/2208.13362}{{\tt arXiv:2208.13362}}.

\bibitem{PhysRevLett.70.3339}
S.~Sachdev and J.~Ye, {\it Gapless spin-fluid ground state in a random quantum
  heisenberg magnet},  {\em Phys. Rev. Lett.} {\bf 70} (May, 1993) 3339--3342,
  [\href{http://arxiv.org/abs/cond-mat/9212030}{{\tt cond-mat/9212030}}].

\bibitem{Maldacena:2016hyu}
J.~Maldacena and D.~Stanford, {\it {Remarks on the Sachdev-Ye-Kitaev model}},
  {\em Phys. Rev. D} {\bf 94} (2016), no.~10 106002,
  [\href{http://arxiv.org/abs/1604.07818}{{\tt arXiv:1604.07818}}].

\bibitem{Maldacena:2016upp}
J.~Maldacena, D.~Stanford, and Z.~Yang, {\it {Conformal symmetry and its
  breaking in two dimensional Nearly Anti-de-Sitter space}},  {\em PTEP} {\bf
  2016} (2016), no.~12 12C104, [\href{http://arxiv.org/abs/1606.01857}{{\tt
  arXiv:1606.01857}}].

\bibitem{Kittu}
A.~Kitaev, ``A simple model of quantum holography (part 1) and (part 2).''
  \url{https://online.kitp.ucsb.edu/online/joint98/kitaev/},
  \url{https://online.kitp.ucsb.edu/online/entangled15/kitaev2/}, 2015.
\newblock Talk given at KITP.

\bibitem{Cotler:2016fpe}
J.~S. Cotler, G.~Gur-Ari, M.~Hanada, J.~Polchinski, P.~Saad, S.~H. Shenker,
  D.~Stanford, A.~Streicher, and M.~Tezuka, {\it {Black Holes and Random
  Matrices}},  {\em JHEP} {\bf 05} (2017) 118,
  [\href{http://arxiv.org/abs/1611.04650}{{\tt arXiv:1611.04650}}]. [Erratum:
  JHEP 09, 002 (2018)].

\bibitem{Maldacena:2015waa}
J.~Maldacena, S.~H. Shenker, and D.~Stanford, {\it {A bound on chaos}},  {\em
  JHEP} {\bf 08} (2016) 106, [\href{http://arxiv.org/abs/1503.01409}{{\tt
  arXiv:1503.01409}}].

\bibitem{Lanczos1950AnIM}
C.~Lanczos, {\it An iteration method for the solution of the eigenvalue problem
  of linear differential and integral operators},  {\em Journal of research of
  the National Bureau of Standards} {\bf 45} (1950) 255--282.

\bibitem{Arnoldipaper}
W.~E. Arnoldi, {\it The principle of minimized iterations in the solution of
  the matrix eigenvalue problem},  {\em Quart. Appl. Math. 9} (1951) 17--29.

\bibitem{Kulkarni:2021xsx}
A.~Kulkarni, T.~Numasawa, and S.~Ryu, {\it {Lindbladian dynamics of the
  Sachdev-Ye-Kitaev model}},  {\em Phys. Rev. B} {\bf 106} (Aug, 2022) 075138,
  [\href{http://arxiv.org/abs/2112.13489}{{\tt arXiv:2112.13489}}].

\bibitem{Sa:2021tdr}
L.~S\'a, P.~Ribeiro, and T.~Prosen, {\it {Lindbladian dissipation of
  strongly-correlated quantum matter}},  {\em Phys. Rev. Res.} {\bf 4} (2022),
  no.~2 L022068, [\href{http://arxiv.org/abs/2112.12109}{{\tt
  arXiv:2112.12109}}].

\bibitem{Garcia-Garcia:2022adg}
A.~M. Garc\'\i{}a-Garc\'\i{}a, L.~S\'a, J.~J.~M. Verbaarschot, and J.~P. Zheng,
  {\it {Keldysh Wormholes and Anomalous Relaxation in the Dissipative
  Sachdev-Ye-Kitaev Model}},  \href{http://arxiv.org/abs/2210.01695}{{\tt
  arXiv:2210.01695}}.

\bibitem{Maldacena:2018lmt}
J.~Maldacena and X.-L. Qi, {\it {Eternal traversable wormhole}},
  \href{http://arxiv.org/abs/1804.00491}{{\tt arXiv:1804.00491}}.

\bibitem{viswanath1994recursion}
V.~Viswanath and G.~M{\"u}ller, {\em The Recursion Method: Application to Many
  Body Dynamics}.
\newblock Lecture Notes in Physics Monographs. Springer Berlin Heidelberg,
  1994.

\bibitem{Kawabata:2022osw}
K.~Kawabata, A.~Kulkarni, J.~Li, T.~Numasawa, and S.~Ryu, {\it {Dynamical
  quantum phase transitions in SYK Lindbladians}},
  \href{http://arxiv.org/abs/2210.04093}{{\tt arXiv:2210.04093}}.

\bibitem{CHOI1975285}
M.-D. Choi, {\it Completely positive linear maps on complex matrices},  {\em
  Linear Algebra and its Applications} {\bf 10} (1975), no.~3 285--290.

\bibitem{JAMIOLKOWSKI1972275}
A.~Jamiołkowski, {\it Linear transformations which preserve trace and positive
  semidefiniteness of operators},  {\em Reports on Mathematical Physics} {\bf
  3} (1972), no.~4 275--278.

\bibitem{Roberts:2018mnp}
D.~A. Roberts, D.~Stanford, and A.~Streicher, {\it {Operator growth in the SYK
  model}},  {\em JHEP} {\bf 06} (2018) 122,
  [\href{http://arxiv.org/abs/1802.02633}{{\tt arXiv:1802.02633}}].

\bibitem{A101280}
D.~Knuth, {\it {The online encyclopedia of integer sequences: A101280}},  2005.

\bibitem{Qi:2018bje}
X.-L. Qi and A.~Streicher, {\it {Quantum Epidemiology: Operator Growth, Thermal
  Effects, and SYK}},  {\em JHEP} {\bf 08} (2019) 012,
  [\href{http://arxiv.org/abs/1810.11958}{{\tt arXiv:1810.11958}}].

\bibitem{Garcia-Garcia:2020ttf}
A.~M. Garc\'\i{}a-Garc\'\i{}a and V.~Godet, {\it {Euclidean wormhole in the
  Sachdev-Ye-Kitaev model}},  {\em Phys. Rev. D} {\bf 103} (2021), no.~4
  046014, [\href{http://arxiv.org/abs/2010.11633}{{\tt arXiv:2010.11633}}].

\bibitem{Susskind:2018tei}
L.~Susskind, {\it {Why do Things Fall?}},
  \href{http://arxiv.org/abs/1802.01198}{{\tt arXiv:1802.01198}}.

\bibitem{Susskind:2020gnl}
L.~Susskind and Y.~Zhao, {\it {Complexity and Momentum}},  {\em JHEP} {\bf 21}
  (2020) 239, [\href{http://arxiv.org/abs/2006.03019}{{\tt arXiv:2006.03019}}].

\bibitem{gruning}
M.~Gr{\"u}ning, A.~Marini, and X.~Gonze, {\it {Implementation and testing of
  Lanczos-based algorithms for random-phase approximation eigenproblems}},
  {\em Computational materials science} {\bf 50} (2011), no.~7 2148--2156,
  [\href{http://arxiv.org/abs/1102.3909}{{\tt arXiv:1102.3909}}].

\bibitem{Garcia-Garcia:2022rtg}
A.~M. Garc\'\i{}a-Garc\'\i{}a, L.~S\'a, and J.~J.~M. Verbaarschot, {\it
  {Universality and its limits in non-Hermitian many-body quantum chaos using
  the Sachdev-Ye-Kitaev model}},  \href{http://arxiv.org/abs/2211.01650}{{\tt
  arXiv:2211.01650}}.

\bibitem{Cai:2022onu}
W.~Cai, S.~Cao, X.-H. Ge, M.~Matsumoto, and S.-J. Sin, {\it {Non-Hermitian
  quantum system generated from two coupled Sachdev-Ye-Kitaev models}},  {\em
  Phys. Rev. D} {\bf 106} (2022), no.~10 106010,
  [\href{http://arxiv.org/abs/2208.10800}{{\tt arXiv:2208.10800}}].

\end{thebibliography}\endgroup

\end{document}